# Easy Access to Bright Oxygen Defects in Biocompatible Single-Walled Carbon Nanotubes *via* a Fenton-Like Reaction

*Simon Settele[1], Florian Stammer[1], Sebastian Lindenthal[1], Simon R. Wald[1], Finn L. Sebastian[1], Han Li[2,3], Benjamin S. Flavel[3] and Jana Zaumseil[1]\**

[1] Institute for Physical Chemistry, Universität Heidelberg, D-69120 Heidelberg, Germany

[2] Department of Mechanical and Materials Engineering, University of Turku, FI-20014 Turku, Finland

[3] Institute of Nanotechnology, Karlsruhe Institute of Technology, Kaiserstraße 12, D-76131 Karlsruhe, Germany

*E-mail: zaumseil@uni-heidelberg.de

# ABSTRACT

The covalent functionalization of single-walled carbon nanotubes (SWNTs) with luminescent oxygen defects increases their brightness and enables their application as optical biosensors or fluorescent probes for *in-vivo* imaging in the second-biological window (NIR-II). However, obtaining luminescent defects with high brightness is challenging with the current functionalization methods due to a restricted window of reaction conditions or the necessity for controlled irradiation with ultraviolet light. Here we report a method for introducing luminescent oxygen defects *via* a Fenton-like reaction that uses benign and inexpensive chemicals without light irradiation. (6,5) SWNTs in aqueous dispersion functionalized with this method show bright $E_{11}$* emission (1105 nm) with 3.2-times higher peak intensities than the pristine $E_{11}$ emission and a reproducible photoluminescence quantum yield of 3%. The functionalization can be performed within a wide range of reaction parameters and even with unsorted nanotube raw material at high concentrations (100 mg $L^{-1}$), giving access to large amounts of brightly luminescent SWNTs. We further find that the introduced oxygen defects rearrange under light irradiation, which gives additional insights into the structure and dynamics of oxygen defects. Finally, the functionalization of ultra-short SWNTs with oxygen defects also enables high photoluminescence quantum yields and their excellent emission properties are retained after surfactant exchange with biocompatible pegylated phospholipids or single-stranded DNA to make them suitable for *in-vivo* NIR-II imaging and dopamine sensing.

# INTRODUCTION

Single-walled carbon nanotubes (SWNTs) can be seen as rolled-up sheets of graphene with a quasi-one-dimensional tubular structure. Their unique optical properties depend on their diameter and roll-up angle, as represented by the index pair (n,m) for each SWNT species or chirality.[1, 2] Semiconducting SWNTs exhibit narrow excitonic emission features ($E_{11}$) in the near-infrared (NIR)[3] that lie well within the second biological window (NIR-II, 1000-1400 nm). In this spectral region, the transparency of most biological tissue is high and light scattering as well as autofluorescence are greatly reduced.[4, 5] This NIR-II emission combined with high photostability and good biocompatibility when noncovalently functionalized with pegylated phospholipids (PL-PEG) or single-stranded DNA (ssDNA) makes SWNTs highly attractive for *in-vivo* fluorescence imaging with high spatial resolution and deep tissue penetration.[6-9] However, despite these advantageous properties of SWNTs, applications are still limited due to their low absolute photoluminescence quantum yield (PLQY) of <1% in aqueous dispersion. The energetically low-lying dark states of SWNTs and the efficient non-radiative decay of highly mobile excitons at quenching sites and nanotube ends limit the achievable PLQY.[10, 11]

A promising tool to increase the absolute PLQY of SWNTs is the introduction of luminescent defects by intentional and controlled covalent functionalization of the $sp^2$ carbon lattice.[12-15] Luminescent defects represent local potential energy wells where mobile $E_{11}$ excitons are trapped, resulting in radiative decay at the defect with lower photon energies and hence emission bands that are red-shifted from the native excitonic $E_{11}$ emission. The localization of excitons at luminescent defects prevents quenching of the fast diffusing excitons at nonradiative defects, leading to overall higher PLQY at optimized defect densities.[16, 17] Over the past decade, a variety of different functionalization methods for the creation of luminescent defects have been developed. Defects that are produced by the formation of a few $sp^3$-hybridized carbon atoms in the SWNT lattice with attached aryl or alkyl groups exhibit bright emission and enable broad synthetic tunability.[16, 18-20] However, the structurally different luminescent oxygen defects have also attracted much attention due their application in *in-vivo* imaging[21, 22] and as single photon emitters.[23]

Luminescent oxygen defects, in which oxygen is incorporated into the SWNT lattice, were first described by Ghosh *et al.* in 2010.[24] They are also referred to as O-defects or oxygen doping.[24, 25] Different oxygen binding configurations were theoretically predicted that should lead to unique changes in the electronic structure of the functionalized SWNTs at the corresponding defect site.[26] The most stable configurations, the ether-d and epoxide-l defects (d-bonds are perpendicular and l-bonds are parallel to the SWNT axis) are expected to result in defect emission that is red-shifted from $E_{11}$ by 135 meV and 310 meV, respectively, for (6,5) SWNTs. The corresponding emission bands are commonly labelled as $E_{11}*$ and $E_{11}*^{-}$.[26] Luminescent oxygen defects can be introduced to SWNTs in aqueous dispersion through reaction with ozone, NaOCl or polyunsaturated fatty acids under ultraviolet (UV) light irradiation.[24, 25, 27] However, the required UV-light makes optimization and upscaling complicated as the duration and intensity of illumination have a large impact on the defect density and final PLQY of the functionalized SWNTs.[28] Thus, a facile method for the controlled creation of luminescent oxygen defects without light irradiation is highly desirable.

One potential option is the well-known Fenton reaction, which was first reported as the decomposition of hydrogen peroxide ($H_2O_2$) with ferrous and ferric ions to form reactive oxygen species (ROS).[29] Since then the name 'Fenton reaction' has been used as an umbrella term for a manifold of related reactions of peroxides with iron ions, but also with copper ions.[30, 31] The Fenton reaction using Fe(II) salts in the presence of $H_2O_2$ was indeed used in the past to covalently functionalize carbon nanotubes, yet harsh reaction conditions and thus extensive functionalization probably prevented the observation of potential luminescent defects.[32, 33]

Here we report an easy and versatile functionalization method that enables the highly controlled and reproducible creation of luminescent oxygen defects in SWNTs in aqueous dispersions by a Fenton-like copper-catalyzed generation of ROS. The produced oxygen-doped SWNTs exhibit a very bright $E_{11}$* emission band with high absolute PLQYs that rival and even surpass reported values for $sp^3$ defects.[17, 34] The method is easily scalable as it does not require light irradiation and only uses inexpensive, readily available, and non-toxic reagents. The defect density can be tuned precisely by various parameters. It can be applied to pre-selected nanotube species such as (6,5) SWNTs coated with different surfactants as well as unsorted small-diameter nanotubes, *e.g.*, CoMoCAT. This simple functionalization method provides access to scalable quantities of bright SWNTs that retain their excellent optical properties even when made biocompatible with pegylated phospholipids (PL-PEG) or single-stranded DNA (ssDNA) and ready to use for high-contrast bio-imaging or bio-sensing in the NIR-II window.

**RESULTS AND DISCUSSION**

**Covalent functionalization *via* Fenton-like reaction**

The creation of luminescent oxygen defects in water-dispersed SWNTs relies on the reaction between *in-situ* generated ROS and the nanotube sidewalls. To achieve and demonstrate this, sodium dodecyl sulfate (SDS)-coated (6,5) SWNTs (referred to as pristine) were obtained by aqueous two-phase extraction (ATPE).[35, 36] They serve as a model system due to their high reactivity and their use in various previous reports on oxygen defects. In our Fenton-like reaction the ROS production is initiated by the reduction of solvated Cu(II) ions (added in the form of $CuSO_4(H_2O)_5$, from here on simply denoted as $CuSO_4$) by an excess of sodium-L-ascorbate (hereafter denoted as NaAsc). We assume the most commonly proposed mechanism for a Fenton-like reaction, which proceeds *via* free radical formation.[31, 37] The reduction of Cu(II) to Cu(I) ions in the presence of dissolved oxygen triggers several oxidation-reduction cycles leading to the formation of $H_2O_2$, which subsequently reacts with Cu(I) to generate hydroxyl (OH·) radicals. These highly reactive hydroxyl radicals attack the SWNT sidewalls resulting in covalent functionalization (see **Figure 1a**).

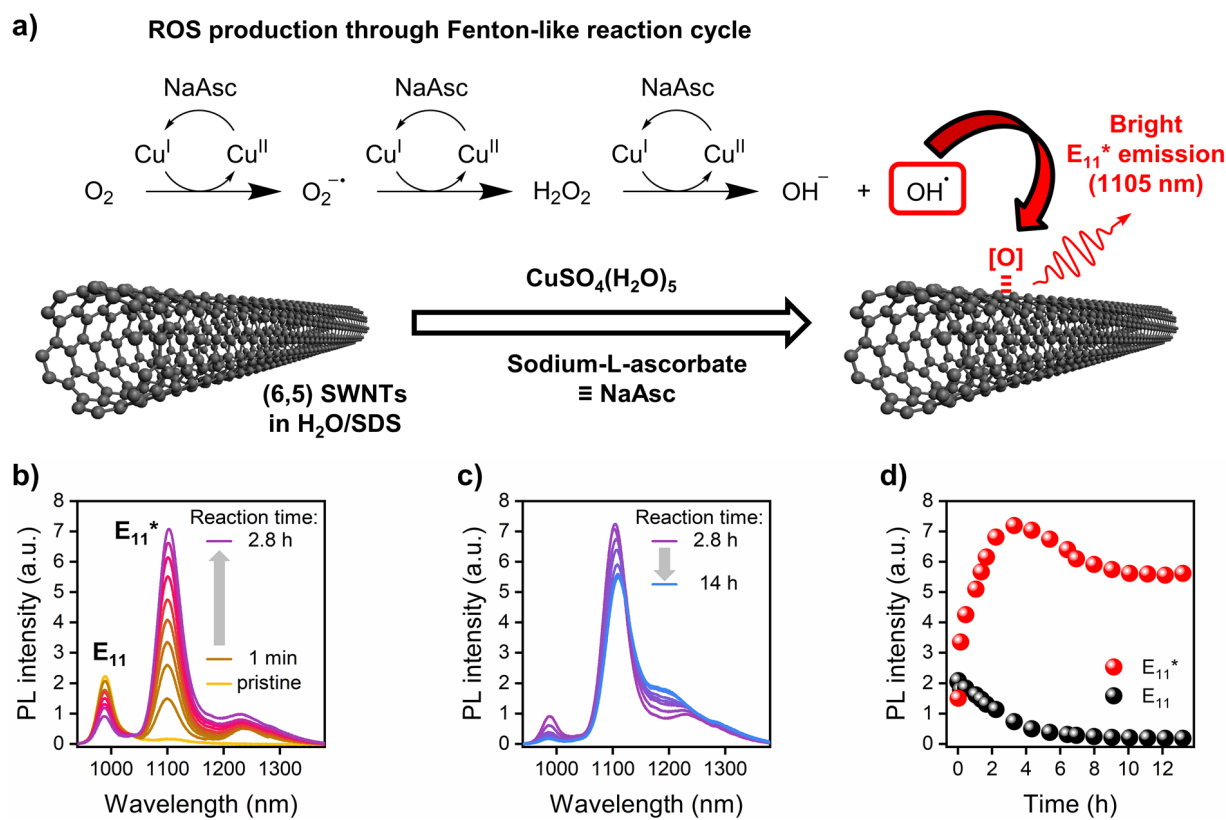

**Figure 1. a)** Reaction scheme for the covalent functionalization of SDS-coated (6,5) SWNTs in water by ROS produced through Fenton-like reaction cycles. The addition of $CuSO_4(H_2O)_5$ and sodium-L-ascorbate (NaAsc) in the presence of dissolved oxygen leads to the formation of hydroxyl radicals and hence luminescent oxygen defects (labelled as $E_{11}^*$). **b)** Evolution of PL spectra of (6,5) SWNTs after addition of 250 μM $CuSO_4(H_2O)_5$ and 3 mM NaAsc for reaction times between 1 min and 2.8 h. **c)** PL spectra after reaction times between 2.8 h and 14 h. **d)** Evolution of absolute $E_{11}$ and $E_{11}^*$ peak intensities with reaction time.

The progress of functionalization of SDS-coated (6,5) SWNTs in dispersion upon addition of $CuSO_4$ (250 μM) and NaAsc (3 mM) can be monitored directly by the evolution of the photoluminescence (PL) spectra (excitation at 570 nm, $E_{22}$) and the appearance of a bright red-shifted emission feature ($E_{11}$*) around ~1105 nm (see **Figure 1b** and **Figure S1a,b** in the Supporting Information). The $E_{11}$* emission feature is red-shifted by 115 nm ($\Delta E$ = 131 meV) relative to the initial $E_{11}$ emission with a full width at half maximum of only 48 nm. Its intensity increases strongly within a few minutes while the $E_{11}$ emission simultaneously decreases. The maximum $E_{11}$* PL intensity is reached after 2.8 hours of reaction time with a 3.2 times higher PL peak intensity (not integrated intensity) than the original $E_{11}$ emission (see **Figure 1b**).

After longer reaction times (> 2.8 h), the $E_{11}$ and $E_{11}$* emission decrease again until the intensity of both emission features reaches a plateau after 14 h (see **Figure 1c** and **1d**). A constant increase of the $E_{11}$*/$E_{11}$ peak intensity ratio is observed until this plateau is reached (see **Figure S1c**). This increased PL ratio can be rationalized by funnelling of mobile $E_{11}$ excitons toward defect sites and is a clear indicator for the introduction of luminescent defects.[16, 18, 20] At very high defect densities, the transport of $E_{11}$ excitons to the luminescent defects is inhibited as the carbon lattice becomes too disrupted, which leads to a decrease in overall PL intensity. In addition to the formation of a bright $E_{11}$* emission band, a second even further red-shifted emission band (~1240 nm) can be observed after short reaction times. However, this emission feature is lower in intensity and appears to show a strong blue-shift with increasing reaction time. The nature of this emission feature and its behavior will be discussed later.

To corroborate our hypothesis that covalent functionalization is initiated by the reduction of Cu(II) species, we performed several reference experiments, in which only NaAsc (3 mM) or only $CuSO_4$ (250 µM) were added to SDS-coated (6,5) SWNTs in water. In both cases no significant changes in the PL spectra were observed (see **Figure S2a,b**). Only a small decrease in $E_{11}$ intensity occurred when $CuSO_4$ was added to the nanotube dispersion, which we attribute to copper-induced PL quenching as previously reported.[38, 39] To ensure that no quenching effects are present after successful functionalization, we added 1.4 mM EDTA (ethylenediaminetetraacetate), a strong metal ion chelating ligand, to bind free copper ions.

Only a small increase in overall PL intensity was detected, thus indicating a negligible effect of copper ions themselves on the defect emission (see **Figure S2c**). Nevertheless, for the reproducible characterization of all functionalized dispersions, EDTA (1.4 mM) was always added to remove residual free copper ions. Subsequently, the reaction mixture was filtered *via* spin-filtration and re-suspended in 1% (w/v) DOC (sodium deoxycholate) to provide a higher dispersion stability for subsequent PL measurements.

**Reaction optimization and optical properties of oxygen defects**

As an alternative to the reaction time, the degree of functionalization can be controlled by the concentration of the reagents when the reaction is allowed to continue in the absence of light for over 16 hours. **Figure 2a** shows that the $E_{11}^*/E_{11}$ PL ratios strongly increase with the $CuSO_4$ concentration for a NaAsc concentration of 12 equivalents (eq.) to $CuSO_4$. The absolute $E_{11}^*$ PL intensity reached its maximum at a concentration of 63 μM $CuSO_4$ (see **Figure S3**) when measured before the final work-up (*i.e.*, addition of EDTA and transfer to DOC).

To unambiguously confirm that the observed $E_{11}^*$ emission is caused by the introduction of defects in the *sp²*-hybridized carbon lattice, we performed absorption and Raman spectroscopy. Absorption spectra (see **Figure S4a**) reveal the presence of an additional absorption feature around ~1105 nm next to the $E_{11}$ absorption band (~987 nm), which can be assigned to the $E_{11}^*$ defect transition. This assignment is confirmed by the linear correlation of the $E_{11}^*/E_{11}$ absorption peak ratios with the $CuSO_4$/NaAsc concentration (shown in **Figure S4b**). More importantly, the introduction of defects is manifested by a continuous increase of the Raman D-mode, as seen in **Figure 2b**, and thus a higher $D/G^+$ mode ratio. $D/G^+$ mode ratio is commonly used as qualitative metric for the measurement of defect densities in SWNTs and has been observed for luminescent oxygen in the past.[17, 24, 25, 40] Again, a linear correlation with the concentration of $CuSO_4$/NaAsc is found (see **Figure S4c**).

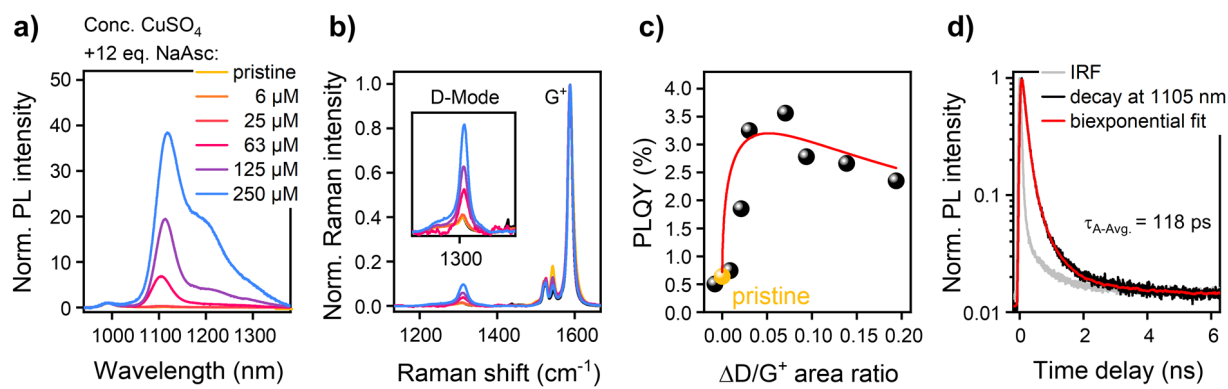

**Figure 2. a)** Normalized (to $E_{11}$) PL spectra of (6,5) SWNTs functionalized after reaction with different concentrations of $CuSO_4$ and 12 eq. of NaAsc and transferred to DOC. **b)** Corresponding normalized (to $G^+$ peak) Raman spectra with a zoom-in on the defect-related D-mode (inset). **c)** Extracted PLQYs of functionalized (6,5) SWNTs (after work-up) *vs.* integrated $\Delta D/G^+$ area ratio as a metric for the defect density (red line is a guide to the eye). The PLQY of pristine (6,5) SWNTs (yellow symbol) represents an average value. **d)** TCSPC histogram of PL decay (black) and biexponential fit (red) of (6,5) SWNTs that were functionalized with 63 μM $CuSO_4$ and 12 eq. NaAsc (after transfer to DOC). The fast $E_{11}$ decay was used as the IRF (grey). An amplitude-averaged lifetime ($\tau_{A\text{-avg.}}$) of 118 ps is obtained.

For an optimized defect density in (6,5) SWNTs produced by the Fenton-like reaction, a ~3.2 times higher $E_{11}^*$ PL peak intensity than the initial $E_{11}$ emission can be achieved. This value is significantly higher than previous reports for oxygen defects.[28] However, similar enhancements in PL intensities have been reported for *sp*[3]-functionalized aqueous dispersions of (6,5) SWNTs.[16] Furthermore, relative increases in PL peak intensity are a poor metric to assess the actual brightness of the final sample and hence its usefulness for applications. Berger *et al.* showed that the brightening factor strongly correlates with the PLQY of the starting material.[18] Hence, we performed absolute PLQY measurements of (6,5) SWNTs that were functionalized at various $CuSO_4$/NaAsc concentrations and subsequently transferred to DOC as shown in **Figure 2c**. Upon functionalization of pristine (6,5) SWNTs (average SWNT length ~500 nm, average PLQY ~0.64% when transferred to DOC) we observe a strong brightening effect with a PLQY of up to 3.6% at optimal defect density. These values correspond to a total brightening factor of ~5.6. All measurements were conducted in $H_2O$ in contrast to the often used $D_2O$. For $D_2O$ dispersions even higher PLQYs are expected as the non-radiative decay of defect state emission by electronic-to-vibrational energy transfer (EVET) would be reduced even further.[41]

Another characteristic of luminescent defects is their long defect-state PL lifetime in comparison to the fast decay of mobile $E_{11}$ excitons. The time-resolved defect-state PL decay was recorded at 1105 nm in a time-correlated single-photon counting (TCSPC) measurement and fitted with a biexponential decay model (see **Figure 2d** and **Table S1**). The extracted amplitude-averaged lifetime ($\tau_{A\text{-avg.}}$) is significantly longer for the $E_{11}*$ emission (118 ps) than for the $E_{11}$ emission (few ps, coinciding with instrument response function, IRF) and agrees with previously reported lifetimes for luminescent oxygen defects and $sp^3$ defects.[23, 41]

One major path for non-radiative decay of trapped excitons is thermally induced de-trapping from the defect, resulting in the temperature-dependent decrease of the $E_{11}*$ emission relative to the $E_{11}$ emission.[42] We also find this effect for the introduced oxygen defects with a thermal trap depth of 97 meV (see **Figure S5**). In addition, only a weak saturation of the $E_{11}*$ emission for higher excitation densities was observed, which mainly results from state-filling of the long-lived defect states (see **Figure S6**). This effect was previously shown to be fairly large for $sp^3$ defects due to their deeper optical trap depths and longer PL lifetimes.[18] A lower dependence of the $E_{11}*/E_{11}$ PL intensity ratios on excitation might be helpful for potential applications in imaging and sensing as it reduces variations when working with different setups and excitations sources.[20]

**Mechanism of functionalization.**

The proposed functionalization mechanism by ROS produced *via* Fenton-like reactions (see above) has several implications that are explored in more detail with the following experiments. Note, no filtration or surfactant transfer to DOC were performed here to provide better insights into the changes of the PL spectra and intensities as reaction conditions were varied. First, the successful functionalization and thus ROS production can only take place in the presence of dissolved oxygen. To confirm this, we greatly reduced the amount of dissolved oxygen by purging the reaction mixture with argon. After addition of CuSO$_4$ and NaAsc no changes in the PL spectra occurred over a 16-hour period (see **Figure 3a**). Thus, the presence of oxygen and the resulting ROS are indeed a necessary requirement for the functionalization process.

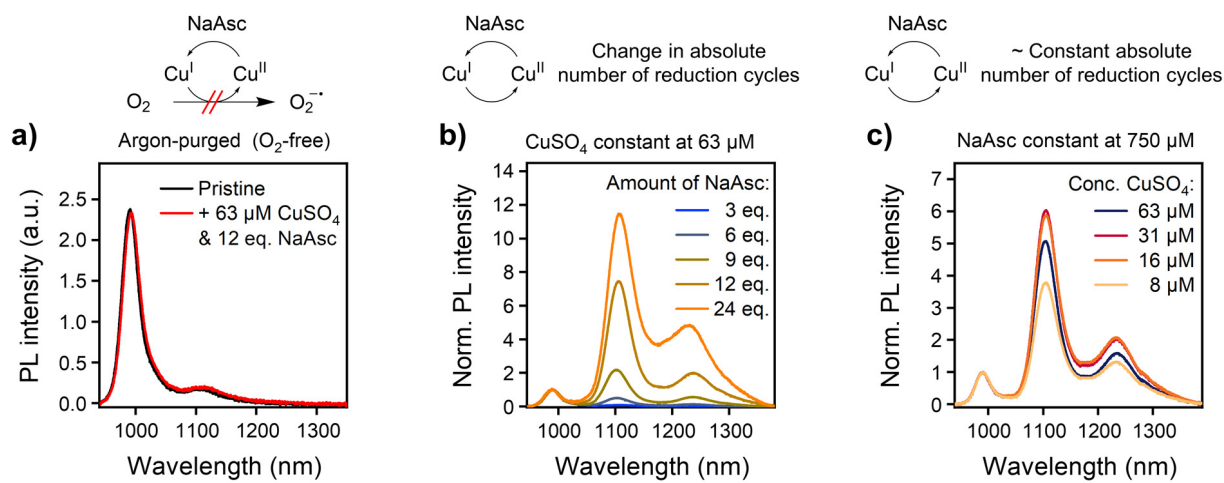

**Figure 3. a)** PL spectra of a dispersion of SDS-coated (6,5) SWNTs in water purged with argon before (black) and after (red) addition of 63 μM CuSO$_4$ and 12 eq. NaAsc. **b)** Normalized PL spectra of SDS-coated (6,5) SWNTs after functionalization with 63 μM CuSO$_4$ and various amounts of NaAsc. **c)** Normalized PL spectra of SDS-coated (6,5) SWNTs after functionalization with various amounts of CuSO$_4$ and 750 μM NaAsc. The reaction time was always 16 h.

The degree of functionalization, represented by the $E_{11}^*/E_{11}$ emission intensity ratio after functionalization, is expected to correlate strongly with the amount of generated ROS species. In the proposed reaction, their quantity is determined by the number of reduction cycles of Cu(II) with the reducing agent NaAsc. Consequently, we find a strong correlation between the degree of functionalization and the NaAsc to Cu(II) ratio. At a constant $CuSO_4$ concentration, almost no functionalization occurs for low ratios (*e.g.*, 3 eq. of NaAsc), while a high degree of functionalization is achieved for high ratios (>12 eq. of NaAsc) as shown in **Figure 3b**. A similar picture arises when the concentration of NaAsc is kept constant and the amount of $CuSO_4$ is decreased. As the absolute number of reduction cycles does not change by reducing the $CuSO_4$ concentration, similar defect densities are obtained (see **Figure 3c** and **Figure S7** for absolute PL spectra). However, this is only the case when NaAsc is present in excess (>12 eq.) and the $CuSO_4$ concentration is sufficiently high. At very low $CuSO_4$ concentrations a decrease in reactivity is observed. This decrease might be related to the disproportionation of Cu(I) species to Cu(0) and Cu(II) species, which is more likely for higher turn-over numbers of the Cu(I) catalyst, and thus at lower Cu(II) concentrations.

The robustness and flexibility of the functionalization procedure with respect to concentrations allows the amount of $CuSO_4$ to be reduced greatly and the defect density to be fine-tuned. A different way to fine-tune the defect density and the $E_{11}^*/E_{11}$ ratio of the functionalized SWNT dispersion is the addition of DOC during the reaction when the desired PL values are reached. The addition of DOC completely inhibits further reactions, as shown in **Figure S8**, due to the dense coverage of the SWNT surface with DOC molecules.[43]

The Fenton-like reaction of Cu(I) with $H_2O_2$ is expected to have a strong dependence on the pH value of the solution in which the functionalization takes place.[30] We performed reactions at different pH values and measured the PL spectra after 16 hours of reaction time. When the functionalization was performed at a pH between 4.8 (standard conditions) and pH 8, no significant changes were observed (see **Figure S9**). At higher pH values the reactivity was reduced probably due to the formation of insoluble $Cu(OH)_2$.

The generation of ROS *via* Fenton-like reactions is not limited to the use of $CuSO_4$ as the metal-catalyst and NaAsc as the reducing agents, but can also be performed with other reagent combinations such as tetraammine copper(II) ions and $Na_2S_2O_4$ (see **Figure S10a**). However, when trying to perform the original Fenton reaction with Fe(II/III) salts and (6,5) SWNTs, we find that Fe(III) ions are unsuitable for functionalization, most likely due to the efficient *p*-doping of SWNTs and thus PL quenching (see **Figure S10b**).[44] Possible alternative reagents for Fenton-like reactions might include the use of other metal salts, such as Co(II).[45]

**Light-induced re-arrangement of defects**

So far it is clear that the reaction of ROS with the carbon lattice of the nanotubes creates luminescent defects but their precise chemical nature and the formation mechanism remain unclear. The optical trap depth of the observed $E_{11}*$ emission (131 meV) agrees very well with previous reports on oxygen defects and corresponding TD-DFT calculations for the most stable ether-d configuration.[26] Throughout this study we observed an additional defect-induced emission peak located around 1240 nm ($E_{11}*^-$), which can be attributed to a second, less stable type of oxygen defect, probably with an epoxide-l configuration. A reaction mechanism for the formation of both types of defects might be proposed by analogy to the atmospheric oxidation of toluene or benzene.[46, 47] One postulated decomposition pathway for toluene and benzene after attack by an OH• radical involves a subsequent reaction with oxygen and the formation of epoxides accompanied be the release of peroxyl radicals. Further rearrangement leads to the formation of, for example, methyl oxepine. A similar reaction mechanism can be formulated for SWNTs and is displayed in **Figure 4a**. Depending on the position of the initial attack of the OH• radical and subsequent thermodynamically favourable rearrangement, the formation of ether-d and epoxide-l defect configurations can be rationalized.[48]

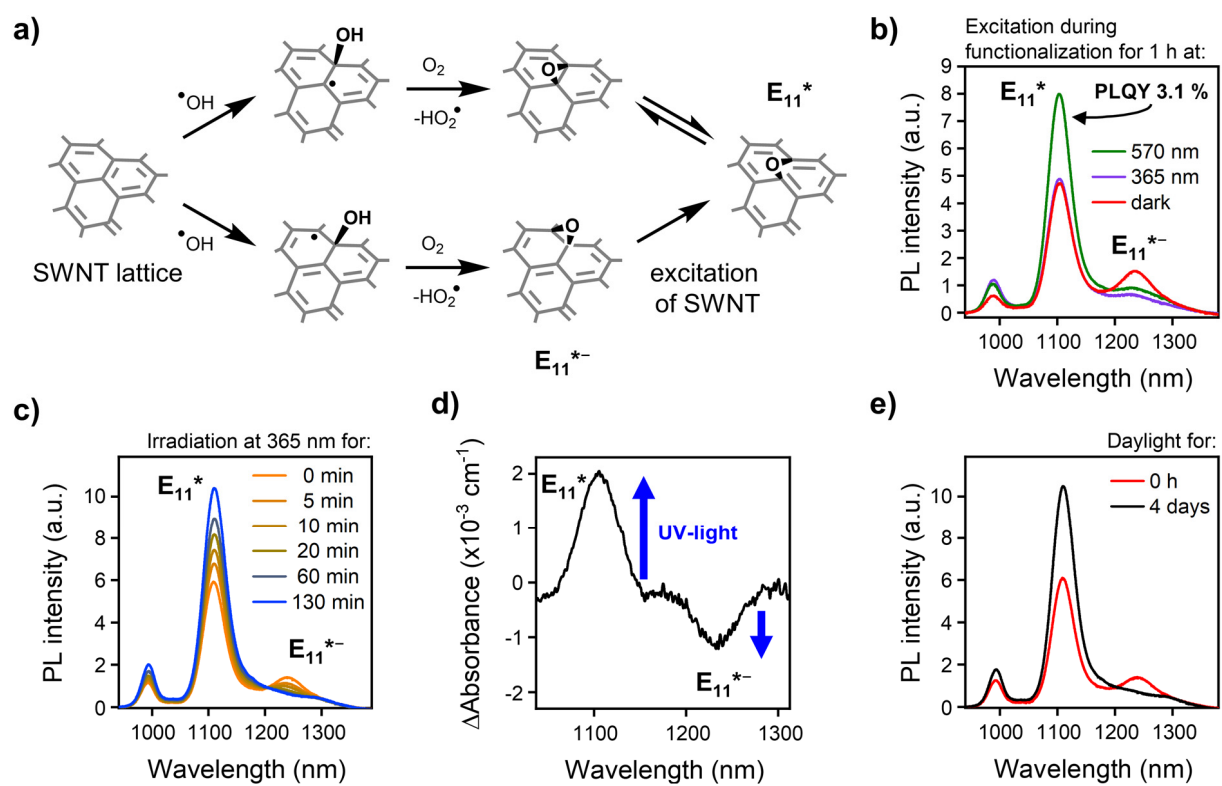

**Figure 4. a)** Proposed functionalization mechanism (simplified) for the formation of different luminescent oxygen defects through reaction with hydroxyl radicals and possible bond rearrangement. **b)** PL spectra after covalent functionalization of SDS-coated (6,5) SWNTs *via* Fenton-like reaction at identical reagent concentrations under $E_{22}$ (570 nm, green) or UV-light (365 nm, purple) excitation or under exclusion of light (red). **c)** PL spectra of functionalized and subsequently DOC-coated (6,5) SWNTs after various UV-light (365 nm) irradiation times. **d)** Differential absorption spectrum of functionalized and DOC-coated (6,5) SWNTs before and after UV-light irradiation (365 nm) for 2 h shows an increase of the absorption band corresponding to the $E_{11}*$ defect while the absorption originating from $E_{11}*^-$ defects decreases. Note, that the absorption cross section for $E_{11}*^-$ defects was estimated to be approximately 3 times smaller than for $E_{11}*$ defects in the case of *sp*$^3$ defects.[17] **e)** PL spectra of functionalized and DOC-coated (6,5) SWNTs before (red) and after (black) exposure to daylight for 4 days.

A striking difference with respect to the emergence of $E_{11}^{*-}$ emission and the overall PL intensity was observed for samples that were exposed to light (*e.g.*, $E_{22}$ excitation, see **Figure 1**) for extended periods of time or kept in the dark (see **Figure 3** or **Figure S3**). To investigate this phenomenon, we performed functionalizations of SDS-coated (6,5) SWNTs with identical reagent concentrations under $E_{22}$ excitation, UV-light irradiation and in the dark with a 1h reaction time (see **Figure 4b**). Clearly, a higher $E_{11}$ (and $E_{11}^*$) intensity and lower $E_{11}^{*-}$ emission are observed for irradiated samples in comparison to the sample that was protected from light. Light irradiation can change the overall functionalization process and thus the higher $E_{11}^*$ PL intensity upon reaction under $E_{22}$ excitation could be explained by an acceleration of the functionalization by optical excitation of the SWNT, as previously observed for the functionalization with diazonium salts.[49] However, light irradiation may also change the optical properties of the functionalized SWNT after the initial reaction is completed.

Consequently, we performed covalent functionalization under exclusion of light and subsequently conducted step-wise UV-light irradiation after the addition of DOC and EDTA to prevent any further reactions with ROS. Surprisingly, we observed a strong increase in $E_{11}$ and $E_{11}^*$ intensity while the $E_{11}^{*-}$ emission was reduced (see **Figure 4c** and **Figure S11**). We can exclude any thermal effects as heating to 80 °C for 0.5 h in the dark had no effect on the PL spectra (see **Figure S12**). Additional Raman measurements and absorption spectroscopy of the functionalized SWNT dispersions before and after UV-light irradiation revealed no change of the $D/G^+$ ratio, but a clear increase in the $E_{11}^*$ absorption band and simultaneous decrease of the $E_{11}^{*-}$ band (see **Figure 4d** and **Figure S13**). This evidence leads us to the conclusion that $E_{11}^{*-}$ defects (epoxide-l configuration) are partially rearranged to form the more stable ether-d defects corresponding to the $E_{11}^*$ emission (see **Figure 4a**). This prominent light-induced reorganization of oxygen defects was proposed previously by Ghosh *et al.* but may have been not observed so far because all prior methods to introduce oxygen defects required light irradiation during the functionalization process.[24, 25] We also considered UV-light and *in-situ* generated $H_2O_2$ as potential causes, yet no significant changes of the PL spectra of pristine (6,5) nanotubes were found (see **Figure S14**) upon addition of $H_2O_2$ or after UV-irradiation.

The reorganization of the defect can even be observed for functionalized nanotubes that are exposed to daylight for more than 4 days (see **Figure 4e**). Samples that were exposed to daylight for extended periods of time, *e.g.* during workup or surfactant exchange, showed similar PL properties even when initially functionalized in the dark (see **Figure 2a**). Consequently, no differences in PLQY are evident when comparing SWNTs functionalized under $E_{22}$ excitation or in the dark (see **Figure S15**).

**Impact of surfactant and SWNT species**

The choice of surfactant has been shown to have a strong influence on the introduction of luminescent oxygen defects due to different binding affinities and densities of the surfactant molecules on the nanotube surface and hence shielding of the SWNT sidewalls.[25, 50] To investigate the effect of various commonly used surfactants, we performed the Fenton-like reaction with different (6,5) SWNT dispersions. **Figure 5a** shows the normalized PL spectra of (6,5) SWNTs after functionalization was performed in 0.33% (w/v) SDS, sodium dodecylbenzene sulfonate (SDBS), sodium cholate (SC) or DOC. While DOC above its critical micelle concentration (CMC) successfully shielded the SWNT from covalent functionalization due to its dense surface coverage, high defect densities could be obtained for SDS and SDBS dispersions (also above CMC). At a concentration of 0.33% (w/v) the CMC of SC was not yet reached, thus leading to incomplete shielding and partial functionalization, while at higher SC concentrations, *e.g.* 0.8% (w/v), no oxygen functionalization was observed (see **Figure S16**). As shown in **Figure S17** (Supporting Information), the highest absolute PL intensities of functionalized (6,5) SWNTs were obtained for SDS-coated nanotube, making SDS the preferred surfactant.

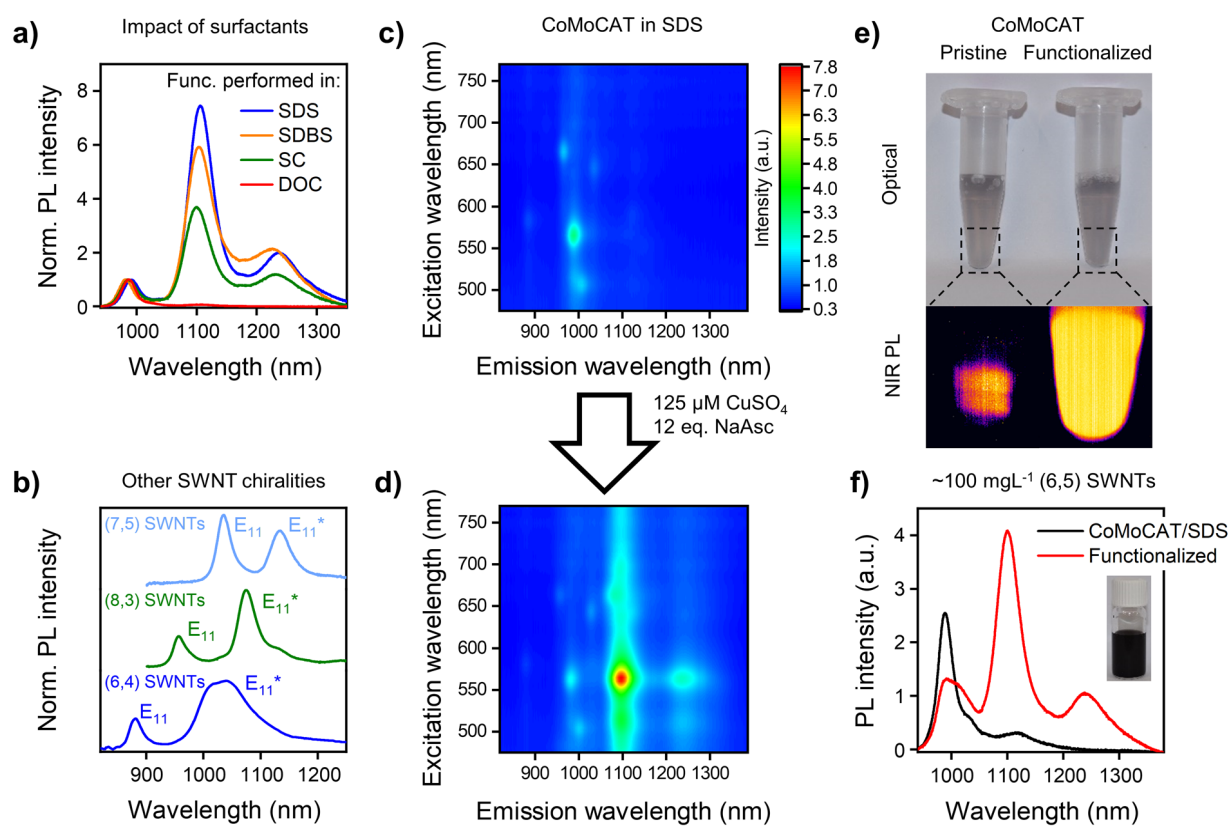

**Figure 5. a)** Normalized PL spectra of (6,5) SWNTs dispersed in SDS (blue), SDBS (orange), SC (green) or DOC (red) (0.33% w/v) after reaction with 63 μM $CuSO_4$ and 12 eq. NaAsc. **b)** Normalized PL spectra of ATPE-sorted and SDS-coated (7,5), (8,3) and (6,4) SWNTs functionalized with 63 μM $CuSO_4$ and 12 eq. NaAsc. **c)** PLE map of unsorted CoMoCAT nanotubes dispersed in SDS and **(d)** PLE map after functionalization of this dispersion with 125 μM $CuSO_4$ and 12 eq. NaAsc. **e)** White light images (top) and NIR (950-1600 nm) PL images (bottom) of a pristine and a functionalized dispersion of CoMoCAT nanotubes. NIR PL images were measured under identical conditions upon excitation at 785 nm at equal concentration. The insets represent the areas that are displayed in the NIR PL images. **f)** PL spectra of diluted unsorted CoMoCAT nanotubes dispersed in aqueous SDS solution before (black) and after (red) functionalization of a highly concentrated dispersion (~100 mg $L^{-1}$ SWNTs).[51]

So far, all reactions were carried out with (6,5) SWNTs. To tune the wavelength of the defect emission band further and demonstrate the applicability of this functionalization method to other nanotube species, we sorted (7,5), (8,3) and (6,4) SWNTs following a modified protocol by Li *et al*.[36] (see **Figure S18, Table S2** and **Experimental Methods**) and functionalized them. We observed chirality-dependent defect emission peaks as shown in **Figure 5b** (for single spectra see **Figure S19**) with smaller optical trap depths for nanotubes with larger diameters (*i.e.*, for (8,3) and (7,5) SWNTs) as expected from previous reports.[16, 20] An increase in overall PLQY was observed for all nanotube species (see **Table S3**), although reactivities decreased with increasing diameter and the defect density was not yet optimized.

While the functionalization of highly purified and sorted SWNT species enabled us to investigate the reaction mechanism and diameter dependence, such purification requires experience and appropriate equipment that may not be readily available to the broader community. To tackle this issue, we performed functionalization of unsorted CoMoCAT raw material dispersed in aqueous SDS solution. This dispersion not only contained several different semiconducting SWNT species (see **Figure S20a**) but also metallic nanotubes. Again, we observed a strong increase in overall PL intensity after functionalization as shown in the PL excitation-emission (PLE) maps in **Figure 5c,d** (for single PL spectra see **Figure S20b**). The strong brightening effect is also clearly visualized by NIR PL images of the corresponding dispersions in Eppendorf cups (see **Figure 5e**, see also **Figure S21** for similar images of sorted (6,5) SWNTs). The striking difference in brightness for the same concentration of nanotubes under identical excitation conditions is evident.

Finally, for many applications and possible commercialization, reproducible large-scale functionalization of nanotubes is highly desirable. The demonstrated Fenton-like reaction for the functionalization of SWNTs can be easily performed at concentrations of unsorted CoMoCat nanotubes of up to ~100 mg L$^{-1}$ because it does not require any irradiation that would be hindered by an opaque dispersion. However, nanotube dispersions at such high concentrations are not very stable in SDS (0.33% w/v) and hence functionalization was performed directly after transferring a stable stock dispersion from DOC to SDS. After successful functionalization with CuSO$_4$ and NaAsc the nanotubes were transferred back to DOC. The concentrations of the highly water-soluble CuSO$_4$ and NaAsc were easily scaled up by the same amount as the SWNT concentration and a strong brightening effect of the CoMoCat dispersions was observed as shown in **Figure 5f**.

**Biocompatible SWNTs with oxygen defects**

For the envisioned application of oxygen-functionalized SWNTs as NIR-II fluorescent labels for *in-vivo* bioimaging or as optical sensors for the detection of, *e.g.* dopamine, they must be made biocompatible without significant reduction of their PLQY. Ultra-short SWNTs (<80 nm) are expected to exhibit lower biotoxicity[52, 53] and hold great promise for high-resolution bioimaging were ultra-small, molecule-like emitter properties are required.[54-56] However, due to quenching of excitons at the nanotube ends their PLQY is almost undetectably low and must be boosted through luminescent defects.[10, 55, 57] To determine the PLQY of very short but oxygen-functionalized SWNTs, we shortened (6,5) SWNTs in SDS by tip-sonication for 4 h yielding an average length of about 53 nm as confirmed by atomic force microscopy (AFM) measurements (**Figure S22**). As shown in **Figure 6a** the emission efficiency of the nanotubes dropped substantially during the sonication process. The shorted SWNTs were subsequently functionalized in the dark as described above and the integrated PL emission increased by a factor of ~7.8 with the E$_{11}$* emission reaching four times the peak intensity of the initial E$_{11}$ emission of the shortened SWNTs. The total PLQY was ~1.1%, which is remarkably high for such short nanotubes and sufficient for *in-vivo* imaging.

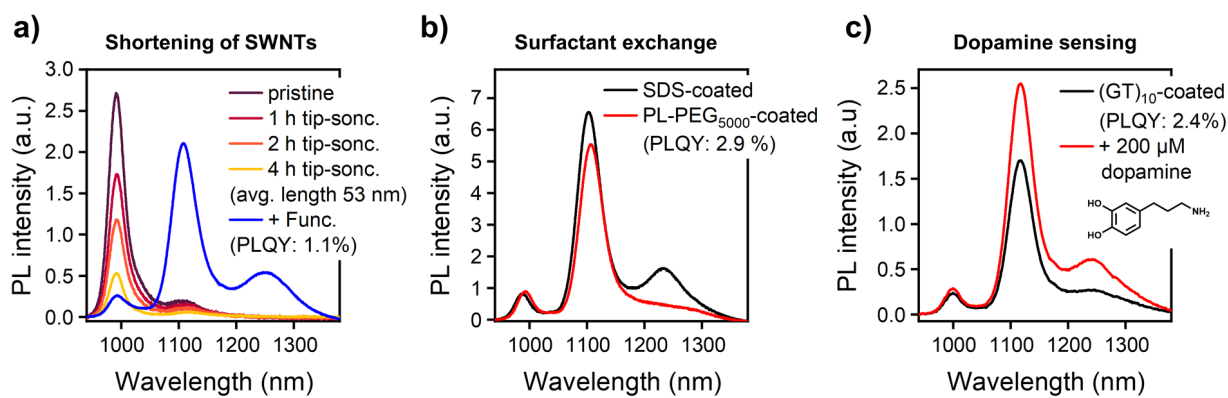

**Figure 6. a)** PL spectra of SDS-coated (6,5) SWNTs after step-wise tip-sonication and subsequent functionalization of the shortened (6,5) SWNTs with an average length of 53 nm. The total PLQY (1.1%) was measured after surfactant transfer to DOC. **b)** PL spectra of functionalized (6,5) SWNTs (not shortened) before and after surfactant transfer from SDS to PL-PEG$_{5000}$ with a final PLQY of 2.9 %. **c)** PL spectra of functionalized (6,5) SWNTs coated with (GT)$_{10}$-ssDNA in PBS buffer before (black, PLQY of 2.4%) and after (red) the addition of 200 µM dopamine.

Two of the most common approaches to provide biocompatibility for SWNTs is non-covalent functionalization with PL-PEG or ssDNA. For this purpose, surfactant transfer to PL-PEG$_{5000}$ and ssDNA-(GT)$_{10}$ in PBS buffer was performed *via* dialysis for 7 and 4 days, respectively.[6, 58] In both cases the successful transfer was confirmed by a red-shift of the $E_{11}$ PL peak position (see **Figure 6b** and **Figure S23**). The emission properties of the functionalized (6,5) SWNTs (not shortened) remained largely unchanged by the transfer and the sample still showed high PLQYs of 2.9% and 2.4%, respectively. Note that during dialysis, the nanotube dispersion was exposed to daylight for several days, and hence the initially present $E_{11}{*}^{-}$ emission vanished as discussed above.

SWNTs wrapped with ssDNA-(GT)$_{10}$ are known for their strong sensitivity towards catecholamines such as dopamine.[59-61] We tested whether ssDNA-(GT)$_{10}$-coated and oxygen-functionalized (6,5) SWNTs still showed a response towards dopamine. In contrast to studies by Spreinat *et al.* for $sp^3$ defects,[62] a strong increase in $E_{11}^*$ intensity occurred upon addition of 200 μM of dopamine to the dispersion as shown in **Figure 6c**. Thus, the presented oxygen-functionalized (6,5) SWNTs could be used for *in-situ* monitoring of dopamine release and are promising candidates for other detection schemes that rely on the interaction of small biomolecules with ssDNA-coated SWNTs.

Finally, we need to address the question of the metal toxicity of SWNTs dispersions that were functionalized in the presence of copper salts. Copper is known for its cytotoxicity at high concentrations and ideally should be removed before conducting *in-vitro* or *in-vivo* experiments.[63] To evaluate whether all copper species can be removed prior to application of SWNT dispersions, we performed covalent functionalization of both (6,5) and CoMoCAT nanotubes followed by washing steps with EDTA as described above. The obtained dispersions had a (6,5) SWNT concentration of ~0.6 mg L$^{-1}$ (based on fitted absorption spectra),[51, 64] which is typical for *in-vivo* measurements.[65] To correctly determine the concentration of the remaining copper species, the SWNTs were digested by addition of $H_2O_2$ and heating for 3 hours at 80 °C until a clear solution was obtained. Absolute copper concentrations were measured by inductively coupled plasma optical emission spectroscopy (ICP-OES). The obtained average copper concentrations for functionalized SWNTs after washing ranged from 51 to 59 μg L$^{-1}$, *i.e.*, approx. 1% of the concentration during functionalization (see **Table S4**). These residual copper concentrations in the functionalized nanotube dispersions are significantly below reported concentrations in human serum (1000-1500 μg L$^{-1}$).[66] In combination with reports on the cell-viability of, *e.g.*, HeLa cells in the presence of Cu(II) ions,[67] we can assume the present copper concentrations to be negligible. Nevertheless, the Cu(II) concentration during functionalization could be further reduced by a factor of 8 while still introducing sufficient oxygen-defects (see **Figure 3c**).

# CONCLUSION

We have developed a very easy, robust and scalable method for the introduction of luminescent oxygen defects in aqueous dispersions of SWNTs using non-toxic, commercially available and inexpensive chemicals. Covalent functionalization of SDS-dispersed (6,5) SWNTs was achieved by continuous and controlled ROS generation through a Fenton-like reaction catalyzed by $Cu^{2+}$ in the presence of the reducing agent sodium ascorbate and dissolved oxygen. Bright defect-induced $E_{11}*$ emission appeared at 1105 nm and a total PLQY of 3% in water was achieved reproducibly. Detailed mechanistic studies revealed that the functionalization conditions can be adapted for the intended application within a large parameter window. In contrast to previous methods, this functionalization procedure of nanotubes with luminescent oxygen defects can be performed under the exclusion of light. Performing the reaction in the dark also revealed a previously unnoticed bond rearrangement of oxygen defects induced by irradiation. The epoxide-l defects ($E_{11}*^-$) with lower stability transform into the more stable ether-d defects ($E_{11}*$) upon optical excitation of SWNTs even by daylight. The presented functionalization method is transferrable to a variety of other SWNT chiralities as well as unsorted CoMoCAT raw material for which a significant brightening effect was observed, even when functionalization was performed on a larger scale and at high SWNT concentrations (~100 mg L$^{-1}$). This scalability enables the production of substantial amounts of bright and biocompatible SWNTs for bioimaging. The oxygen-functionalized SWNTs retain their high PLQY after coating with PL-PEG$_{5000}$ and show clear responsiveness to dopamine after wrapping with ssDNA-(GT)$_{10}$. Even ultra-short (<100 nm) functionalized SWNTs still show sufficiently high PLQY in the second biological window for bioimaging. In conclusion, Fenton-like reactions using copper salts and sodium ascorbate are ideal to functionalize SWNTs with luminescent oxygen defects that enable high PLQY in the near-infrared under a wide range of conditions and for many applications.

# EXPERIMENTAL METHODS

**Selective Dispersion of (6,5) SWNTs.** The selection of aqueous dispersed (6,5) SWNTs from CoMoCAT raw material (CHASM SG65i-L58) was performed by aqueous two-phase extraction (ATPE) as described previously.[17] Raw material dispersed in deoxycholate (DOC, BioXtra) was mixed in a two-phase system composed of dextran ($M_w$ = 70 kDa, TCI) and poly(ethylene glycol) (PEG, $M_w$ = 6 kDa, Alfa Aesar). (6,5) SWNTs were separated following a diameter sorting protocol by addition of sodium dodecyl sulfate (SDS, Sigma-Aldrich). Metallic and semiconducting SWNTs were separated by addition of sodium cholate (SC, Sigma-Aldrich) and sodium hypochlorite (NaClO, Sigma-Aldrich). The selected (6,5) SWNTs in DOC were transferred to SDS by concentrating them in a pressurized ultrafiltration stirred cell (Millipore) with a 300 kDa $M_w$ cutoff membrane and sub-sequent addition of 1% (w/v) SDS.

**Selective Dispersion of (6,4), (8,3) and (7,5) SWNTs.** Separation of (6,4), (8,3) and (7,5) SWNTs was performed following a modified protocol by Li *et al.*[36] CoMoCAT raw material (Sigma Aldrich, batch MKCR4865) dispersed in deoxycholate (DOC, BioXtra) was mixed in a two-phase system composed of dextran ($M_w$ = 70 kDa, TCI) and poly(ethylene glycol) (PEG, $M_w$ = 6 kDa, Sigma-Aldrich) and 0.5% (w/v) SDS (Sigma-Aldrich) with a total volume of 4 mL. The concentration of DOC was adjusted by addition of the appropriate amount of CoMoCAT raw material dispersed in DOC (1% w/v). Separation of (6,4) SWNTs was achieved at 0.025% (w/v) DOC and additional 0.5% (w/v) SC. Separation of (8,3) and (7,5) SWNTs was accomplished at 0.05% (w/v) and 0.08% (w/v) DOC, respectively. Mimic suspensions were prepared in the same way without dispersed CoMoCAT raw material and DOC was added to adjust the desired DOC concentration. SWNT separation was performed in a three-steps procedure by addition of HCl and separation of metallic and semiconducting SWNTs was aided by further addition of NaOCl (Roth). Exact volumes of added HCl and NaOCl are given in

**Table S2**. ATPE-polymers were removed *via* multiple spinfiltration steps (Amicon Ultra-4, 100 kDa) and purified SWNTs were resuspended in 0.33% (w/v) SDS.

**Shortening of (6,5) SWNTs.** A dispersion of (6,5) SWNTs in 1% (w/v) SDS obtained by ATPE was adjusted to an optical density of 0.33 cm$^{-1}$ at the $E_{11}$ transition with ultra-pure water and tip-sonicated stepwise for 4 h (Sonic Vibra Cell VCX500) while immersed in a 5 °C water cooling bath. The sonicated dispersions were centrifuged at 60,000 g for 30 min (Beckman Coulter Avanti J-26S XP), and the supernatant was used for functionalization.

**Introduction of Luminescent Defects *via* Fenton-like Reaction.** For the introduction of luminescent defects *via* Fenton-like reaction, the optical density of the aqueous SWNT dispersion was adjusted to 0.33 cm$^{-1}$ at the $E_{11}$ transition for (6,5) and (6,4) SWNTs or 0.2 cm$^{-1}$ at the $E_{11}$ transition for (7,5) and (8,3) SWNTs with ultra-pure water or 0.33% (w/v) SDS. For functionalization of unsorted SWNTs, CoMoCAT raw material dispersed in 1% (w/v) DOC was transferred to 0.33% (w/v) SDS prior to functionalization and adjusted to 0.33 cm$^{-1}$ at $E_{11}$ transition of (6,5) SWNTs. Functionalization was performed directly after transfer to SDS.

Fresh stock solutions of $CuSO_4(H_2O)_5$ (Sigma-Aldrich, ≥99.9% trace-metal basis) and sodium-L-ascorbate (Sigma-Aldrich, ≥99%) were prepared with a typical concentration of 6.25 mg mL$^{-1}$ and 40 mg mL$^{-1}$, respectively. Aliquots of $CuSO_4(H_2O)_5$ and sodium ascorbate stock solutions were added to the reaction mixture until the desired concentrations were reached and the reaction mixture was stored in the dark for 16 h. For functionalization under light excitation, the reaction mixture was placed either in a Fluorolog-3 spectrometer (Horiba Jobin-Yvon, see above) for excitation at the $E_{22}$ transition (570 nm) or irradiated with UV-light (365 nm, SOLIS-365C, Thorlabs, 1.9 mW mm$^{-2}$). To stop the reaction, aliquots of a 1.4 M Na$_4$EDTA (10 µL per 1 mL reaction volume, Sigma-Aldrich, 98%) solution and 10% (w/v) DOC solution (20 µL per 1 mL reaction volume) were added to the reaction mixture. For reorganization of oxygen defects the functionalized SWNT dispersions may be irradiated with UV-light (365 nm,

SOLIS-365C, Thorlabs, 1.9 mW mm$^{-2}$) for 2 h or placed in daylight for >4 days. Finally, the reaction mixture was filtered *via* spin-filtration (Amicon Ultra-4, 100 kDa) and functionalized (6,5) SWNTs were resuspended in 1 % (w/v) DOC.

**Surfactant Exchange to PL-PEG$_{5000}$.** Functionalized (6,5) SWNTs were transferred to PL-PEG$_{5000}$ according to an adapted procedure by Welsher *et al.*[6] Functionalized (6,5) SWNTs were mixed with an appropriate amount of 18:0 PEG5000PE (1,2-distearoyl-*sn*-glycero-3-phosphoethanolamine-*N*-[methoxy(polyethylene glycol)-5000, PL-PEG$_{5000}$], Avanti Lipids) to achieve a final PL-PEG$_{5000}$ concentration of 2 mg mL$^{-1}$. The mixture was transferred to a 1 kDa dialysis bag (Spectra/Por®, Spectrum Laboratories Inc.) and dialyzed for 7 days against ultra-pure water. Lastly, the obtained dispersion was bath sonicated for 15 minutes.

**Surfactant Exchange to ssDNA.** Functionalized (6,5) SWNTs coated with ssDNA were obtained following a protocol by Ackermann *et al.*[58] First, the functionalized (6,5) SWNTs were concentrated to an absorbance of 2.0 at the $E_{11}$ transition *via* spin-filtration (Amicon Ultra-4, 100 kDa). The concentrated functionalized (6,5) SWNTs (1 mL) were mixed with ssDNA (100 µL of (GT)$_{10}$, 2 mg mL$^{-1}$ in phosphate buffered saline solution (PBS, Carl Roth) and transferred to a 1 kDa dialysis bag (Spectra/Por®, Spectrum Laboratories Inc.). After dialysis for 4 days against PBS the SWNT dispersion was centrifuged for 20 min at 20,000 g (Hettich Micro 220R) and the supernatant was used for characterization.

**Basic SWNT Characterization.** Baseline-corrected absorption spectra were acquired with a Cary 6000i UV-VIS-NIR spectrophotometer (Varian, Inc.). Resonant Raman spectroscopy was performed with a Renishaw inVia Reflex confocal Raman microscope, equipped with a 50× long-working distance objective (Olympus, N.A. 0.5). The SWNT dispersions were dropcast onto glass slides and subsequently rinsed with water to remove any excess surfactants. Raman spectra were acquired under 532 nm laser excitation. Atomic force microscope (AFM) images of shortened functionalized (6,5) SWNTs were recorded with a Bruker Dimension Icon AFM in tapping mode under ambient conditions. For the statistical analysis of the length of individual SWNTs, Si/SiO$_2$ substrates were incubated with aqueous poly-L-lysine hydrobromide (PLL, 0.1 g L$^{-1}$) for 10 min, rinsed with water and blow-dried with nitrogen. A droplet of a diluted SWNT dispersion of functionalized shortened (6,5) SWNTs was then placed on the PLL treated Si/SiO$_2$ substrate and incubated for 10 min. The droplet was carefully removed and the substrate gently washed with ultra-pure water and subsequently blow-dried with nitrogen.

**Photoluminescence Spectroscopy.** Photoluminescence (PL) spectra and emission-excitation (PLE) maps were obtained with a Fluorolog-3 spectrometer (Horiba Jobin-Yvon) equipped with 450 W xenon arc-discharge lamp (450 W) and a liquid nitrogen cooled InGaAs line camera (Symphony II). For temperature dependent PL measurements between 293 K and 353 K, a Peltier-based temperature-controlled cuvette holder was used. For acquisition of PL spectra at high excitation densities, for PL lifetime measurements and for the determination of photoluminescence quantum yield (PLQY) values, a home-built setup was used. Briefly, functionalized (6,5) SWNTs were excited at 570 nm with the wavelength-filtered output of a pico-second pulsed supercontinuum laser (NKT Photonics SuperK Extreme) focused by a 50× NIR-optimized objective (N.A. 0.65, Olympus). Residual laser light was blocked by a dichroic long-pass filter (875 nm cut-off) and additional long-pass filter (cut-on 830 nm). PL spectra were recorded with an Acton SpectraPro SP2358 spectrograph (grating blaze 1200 nm, 150 lines mm$^{-1}$) equipped with a liquid-nitrogen-cooled InGaAs line camera (Princeton Instruments, OMA-V:1024).

**Lifetime Measurements.** For wavelength-dependent PL lifetime measurements in a time-correlated single photon counting (TCSPC) scheme, the output of the spectrograph was guided onto a gated InGaAs/InP avalanche photodiode (Micro Photon Devices) *via* an NIR-optimized 20× objective (Mitutoyo, N.A. 0.40). Histograms of photon arrival times were recorded with a counting module (PicoQuant PicoHarp 300) and reconvolution-fitted using the SymPhoTime 64 software. The short decay of $E_{11}$ excitons in a (6,5) SWNT network served as the instrument response function.

**PL Quantum Yield Measurements.** Absolute PLQYs ($\eta$) of pristine and functionalized dispersions were determined from integrating sphere measurements, which give direct access to the ratio of emitted ($N_{em}$) and absorbed photons ($N_{abs}$) as reported previously:[11, 18]

$$\eta = \frac{N_{em}}{N_{abs}} \quad (1)$$

For these measurements, the SWNT dispersions were adjusted to an optical density of <0.2 cm$^{-1}$ at the $E_{11}$ transition. A quartz glass cuvette filled with 1 mL of diluted SWNT dispersion was placed in the center of an integrating sphere (LabSphere, Spectralon coating). After excitation at the respective $E_{22}$ transition the light exciting the integrating sphere (attenuated laser light and photoluminescence) was guided to the spectrometer *via* an optical fiber. The same measurement was repeated with the pure solvent (*e.g.* 1% (w/v) DOC) to account for light absorption/scattering by the solvent. A value proportional to the number of emitted photons was determined by integration of the sample emission spectrum. The number of absorbed photons was calculated from the difference of the integrated laser signals of the reference (pure solvent) and measurement sample (SWNT dispersion). The wavelength-dependent detection efficiency and other optical losses were accounted for by collection of calibration spectra using a stabilized tungsten halogen light source with known spectral power distribution (Thorlabs SLS201/M,300-2600 nm).

**NIR-PL Imaging.** NIR-PL images (950 – 1600 nm) of SWNT dispersions were obtained by excitation with an expanded beam of a 785 nm laser diode (Alphalas Picopower-LD-785-50) operated in continuous wave mode. The emitted light was focused with a tube lens on a thermoelectrically cooled In-GaAs camera (Xenics XEVA-CL-TE3, 252x320 pixels). A 950 nm long-pass filter was used to filter out residual laser light. Pristine and functionalized SWNT dispersions were imaged under identical conditions.

**Determination of Copper Concentration with ICP-OES.** To prepare samples for inductively coupled plasma optical emission spectroscopy (ICP-OES) measurements, ATPE-selected (6,5) SWNTs with a total volume of 4 mL were functionalized *via* Fenton-like reaction and processed as described above. Functionalization was performed at a $CuSO_4(H_2O)_5$ concentration of 63 µM and 750 µM sodium-L-ascorbate. The same functionalization was applied to unsorted CoMoCAT SWNTs. To ensure efficient removal of copper, the reaction mixtures were filtered four times *via* spin-filtration (Amicon Ultra-4, 100 kDa) and re-suspended in 1 mL 0.1% (w/v) DOC. For total digestion of SWNTs, the obtained dispersion was mixed with 1 mL $H_2O_2$ (30%, Sigma Aldrich) and heated to 80 °C while being magnetically stirred until the solution became transparent. The obtained solution was diluted with 2 mL ultra-pure water, acidified with 50 µL nitric acid (65%, Suprapur) and filtered (polytetrafluoroethylene (PTFE) syringe filter, 5 µm pore size). Identical processing, without the addition of $CuSO_4(H_2O)_5$ and sodium-L-ascorbate was performed for pristine ATPE-selected (6,5) SWNTs and pristine unselected CoMoCAT SWNTs. An additional reference sample was pre-pared without the addition of SWNTs. All samples were prepared in duplicate. The absolute copper concentrations of these samples were determined by ICP-OES (Agilent ICP-OES 720) at λ = 327.395 nm. A copper single element standard solution from Ultra Scientific (1000 mg $L^{-1}$) was used to create a calibration curve.

## AUTHOR INFORMATION

**Corresponding Author**

*E-mail: zaumseil@uni-heidelberg.de

**Author Contributions**

S.S. and F.S. fabricated all samples and S.S. analyzed the data. S.L. contributed to AFM and Raman measurements. S.R.W. performed ATPE-sorting of (6,4), (8,3) and (7,5) SWNTs. F.L.S. and S.S performed NIR-PL imaging. H.L. and B.S.F. provided ATPE-sorted (6,5) SWNTs. J.Z. conceived and supervised the project. S.S. and J.Z. wrote the manuscript. All authors discussed the data analysis and commented on the manuscript.

**Notes**

The authors declare the following competing financial interest(s): S.S., F.S. and J.Z. are listed as inventors on a patent application describing the functionalization approach presented in this study.

## ACKNOWLEDGMENT

This project has received funding from the European Research Council (ERC) under the European Union's Horizon 2020 research and innovation programme (Grant agreement No. 817494 "TRIFECTs"). B.S.F. and H.L. gratefully acknowledge support by the DFG under grant numbers FL 834/5-1, FL 834/7-1, FL 834/9-1 and FL 834/12-1. H.L. acknowledges financial support from the Turku Collegium for Science, Medicine and Technology (TCSMT). S.S. thanks C. Alexander Schrage for valuable input on dopamine sensing with ssDNA-wrapped SWNTs.
29

SUPPORTING INFORMATION

# Easy Access to Bright Oxygen Defects in Biocompatible Single-Walled Carbon Nanotubes *via* a Fenton-Like Reaction

*Simon Settele[1], Florian Stammer[1], Finn L. Sebastian[1], Sebastian Lindenthal[1], Simon R. Wald[1], Han Li[2,3], Benjamin S. Flavel[3] and Jana Zaumseil[1]\**

[1] Institute for Physical Chemistry, Universität Heidelberg, D-69120 Heidelberg, Germany

[2] Department of Mechanical and Materials Engineering, University of Turku, FI-20014 Turku, Finland

[3] Institute of Nanotechnology, Karlsruhe Institute of Technology, Kaiserstrasse 12, D-76131 Karlsruhe, Germany

\* corresponding author: *zaumseil@uni-heidelberg.de*

# Contents

**Characterization of functionalized (6,5) SWNTs**

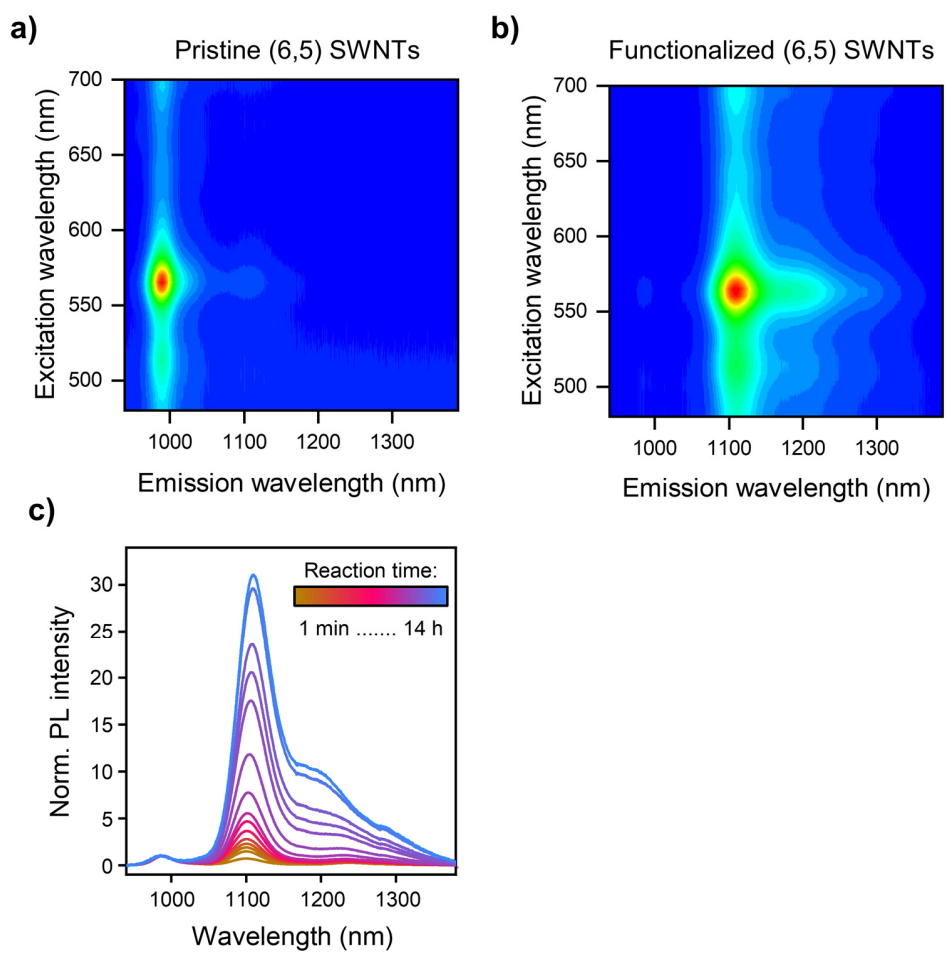

**Figure S1**. Spectral evolution of (6,5) SWNTs functionalized *via* Fenton-like reaction. a) PLE map of ATPE-selected (6,5) SWNTs in aqueous 0.33% (w/v) SDS dispersion. b) PLE map of functionalized (6,5) SWNTs after addition of 250 µM $CuSO_4$ and 3 mM NaAsc and 14 h reaction while excited at the $E_{22}$ transition. c) Normalized (to $E_{11}$) PL spectra measured after reaction times between 1 min and 14 h.

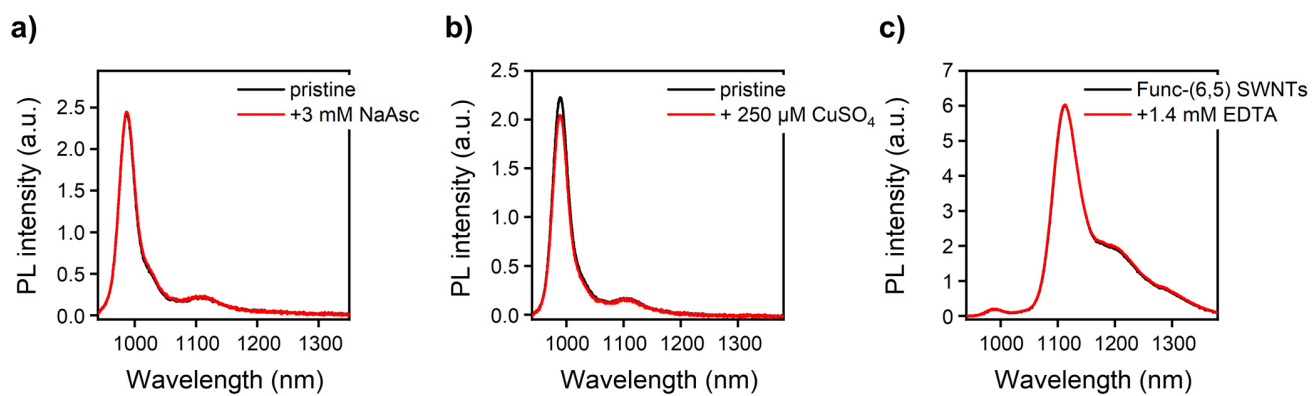

**Figure S2**. Reference experiments. PL spectra of SDS-coated (6,5) SWNTs before (black) and after (red) the addition of a) 3 mM NaAsc or b) 250 µM CuSO4. c) PL spectra of functionalized (6,5) SWNTs in SDS before and after addition of 1.4 mM EDTA to remove residual free copper ions.

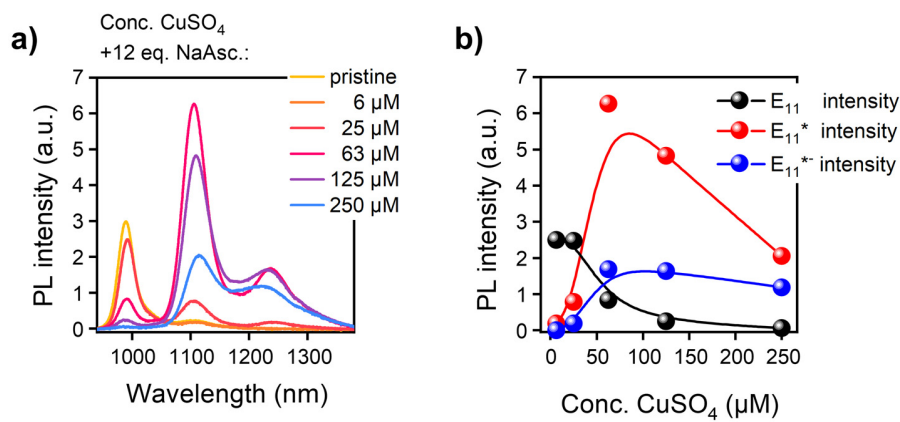

**Figure S3.** Concentration-dependend covalent functionalization of SDS-coated (6,5) SWNTs. a) PL spectra after functionalization at various concentrations of CuSO4 and 12 eq. NaAsc. b) Evolution of total PL intensities ($E_{11}$, $E_{11}^*$, $E_{11}^{*-}$) after functionalization with various concentrations of CuSO4 and 12 eq. NaAsc.

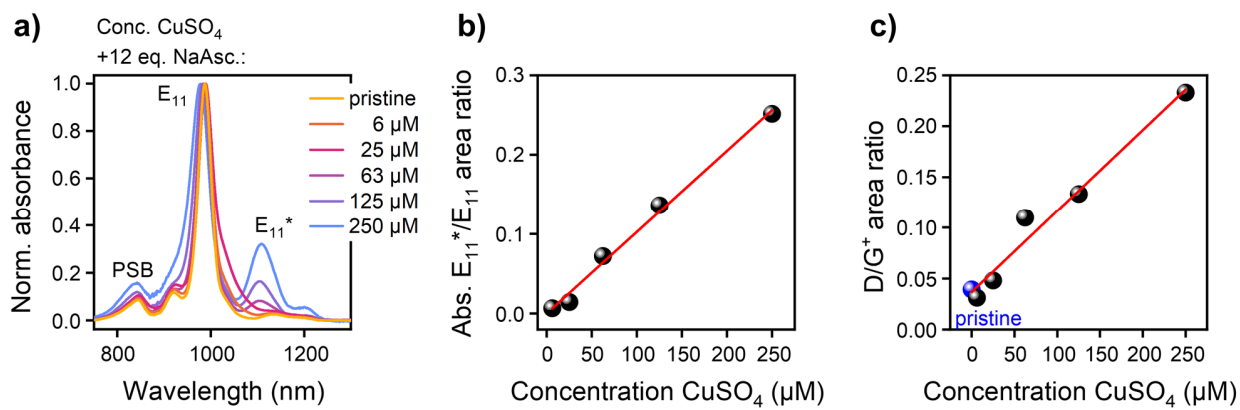

**Figure S4.** Defect density metrics. a) Normalized (to $E_{11}$) absorption spectra of (6,5) SWNTs functionalized *via* Fenton-like reaction at various concentrations of $CuSO_4$ and 12 eq. NaAsc. b) Extracted absorption $E_{11}^*/E_{11}$ area ratio *vs.* concentration of $CuSO_4$. c) Raman $D/G^+$ area ratio extracted from Raman measurements *vs.* concentration of $CuSO_4$. Red lines are linear fits to the data.

**Table S1.** Extracted short and long lifetime components ($\tau_{short}$, $\tau_{long}$) and corresponding normalized amplitudes ($A_{short}$, $A_{long}$) for defect emission from (6,5) SWNTs functionalized *via* Fenton-like reaction (with 62.5 µM $CuSO_4$) and subsequently dispersed in aqueous DOC solution. An amplitude-averaged lifetime ($\tau_{A\text{-avg.}}$) of 118 ps is obtained.

| Sample | $\tau_{long}$ | $\tau_{short}$ | $A_{long}$ | $A_{short}$ | $\tau_{A\text{-avg.}}$ |
|---|---|---|---|---|---|
| Func. (6,5) SWNTs (measured in DOC at 1105 nm) | 220 ps | 86 ps | 0.24 | 0.76 | 118 ps |

**Temperature dependence of defect state photoluminescence**

Trapping of excitons at defect sites is expected to be reversible when the thermal energy $kT$ is larger than the potential well depth of the defect site and thus the detrapping energy $\Delta E_{\text{Thermal}}$. According to Kim *et al.*, $\Delta E_{\text{Thermal}}$ can be determined from PL measurements at different temperatures ($T$) and construction of a van't Hoff plot:[1]

$$\frac{I_{E_{11}}}{I_{E_{11}^*}} \propto e^{-\left(\frac{\Delta E_{\text{Thermal}}}{kT}\right)} \quad (1)$$

$$\ln\left(\frac{I_{E_{11}}}{I_{E_{11}^*}}\right) = -\left(\frac{\Delta E_{\text{Thermal}}}{kT}\right) + A \quad (2)$$

where $I_{E_{11}}$ and $I_{E_{11}^*}$ are the integrated PL intensities, $k$ the Boltzmann constant and $A$ a correction factor. We performed PL measurements at temperatures between 283 K and 323 K (see **Figure S5a**). The extracted integrated PL intensities were fitted with a linear regression model. According to this analysis, the E$_{11}$* defect exhibits $\Delta E_{\text{Thermal}}$ of approximately 96.7 meV (see **Figure S5b**), which is similar to previously reported values for oxygen defects.[2]

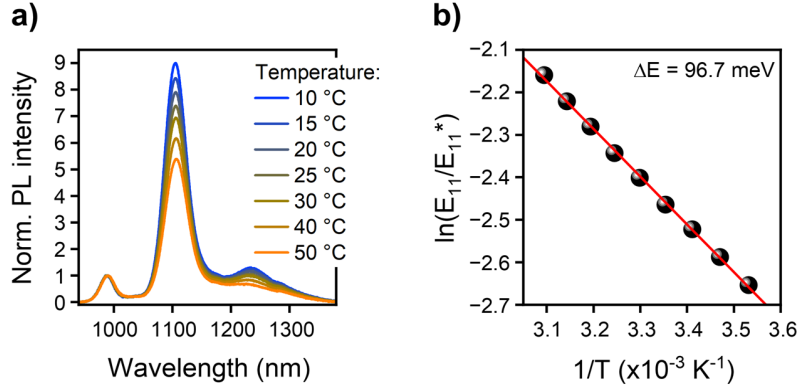

**Figure S5.** Temperature dependent defect state PL. a) Temperature-dependent, normalized PL spectra of a dispersion of DOC-coated (6,5) SWNTs functionalized *via* Fenton-like reaction. b) van't Hoff plot for the E$_{11}$* emission band and linear fit to the data. A detrapping energy ($\Delta E_{\text{Thermal}}$) of 96.7 meV was extracted.

**Impact of excitation density**

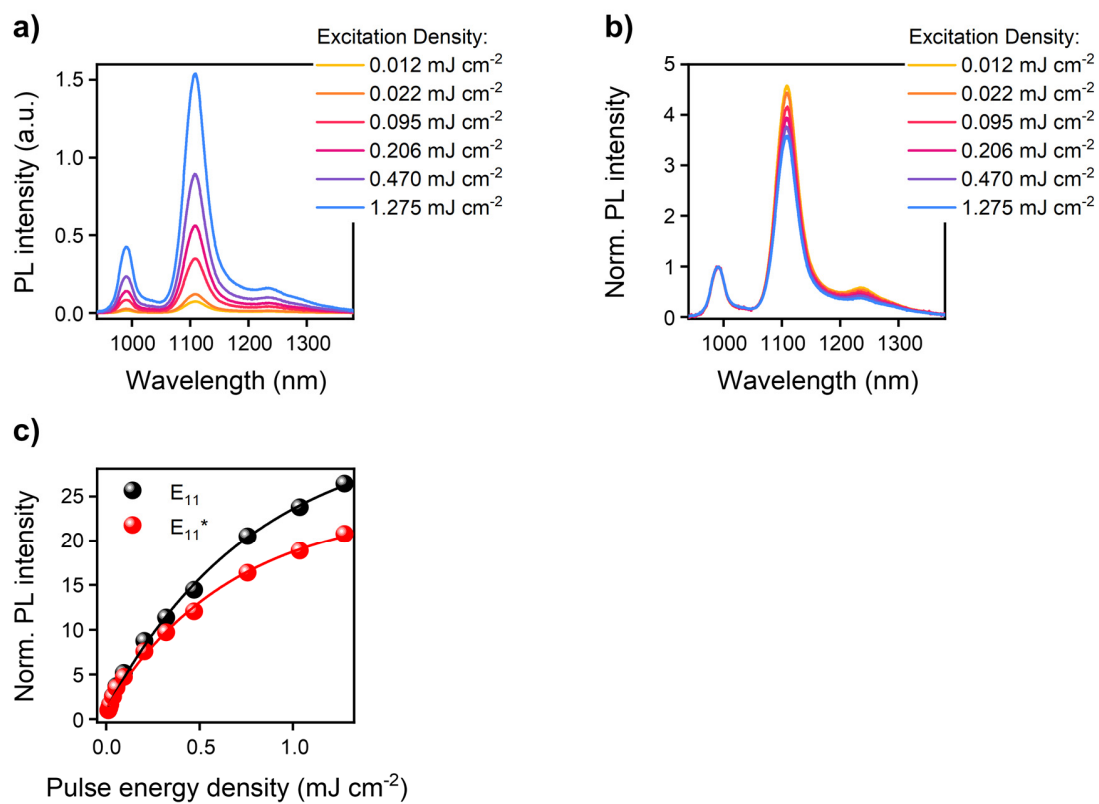

**Figure S6.** Excitation power-dependence of $E_{11}^*$ defect state PL. Absolute PL (a) and normalized (b) PL spectra of (6,5) SWNTs functionalized *via* Fenton-like reaction recorded at different excitation densities (pulsed excitation at $E_{22}$). c) $E_{11}$ and $E_{11}^*$ PL intensities normalized to their intensity at lowest excitation density *vs.* pulse energy density (lines are shown as guides to the eye).

**Impact of various reaction parameters**

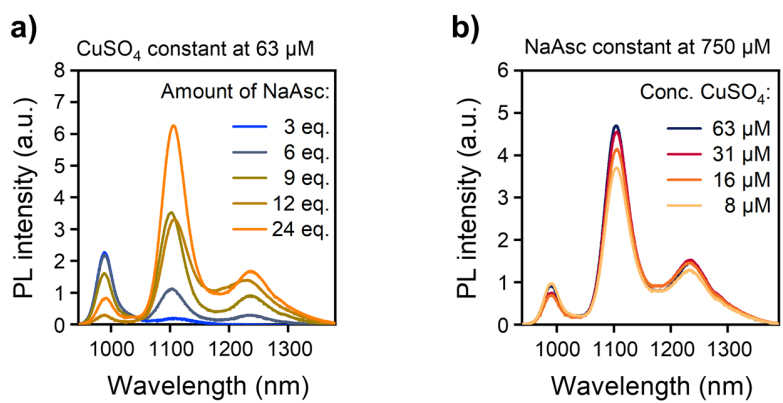

**Figure S7.** Effect of reagent concentrations. a) PL spectra of SDS-coated (6,5) SWNTs functionalized *via* Fenton-like reaction at constant $CuSO_4$ concentration and various concentrations of NaAsc. b) PL spectra of SDS-coated (6,5) SWNTs functionalized *via* Fenton-like reaction at constant NaAsc concentration and various concentrations of $CuSO_4$.

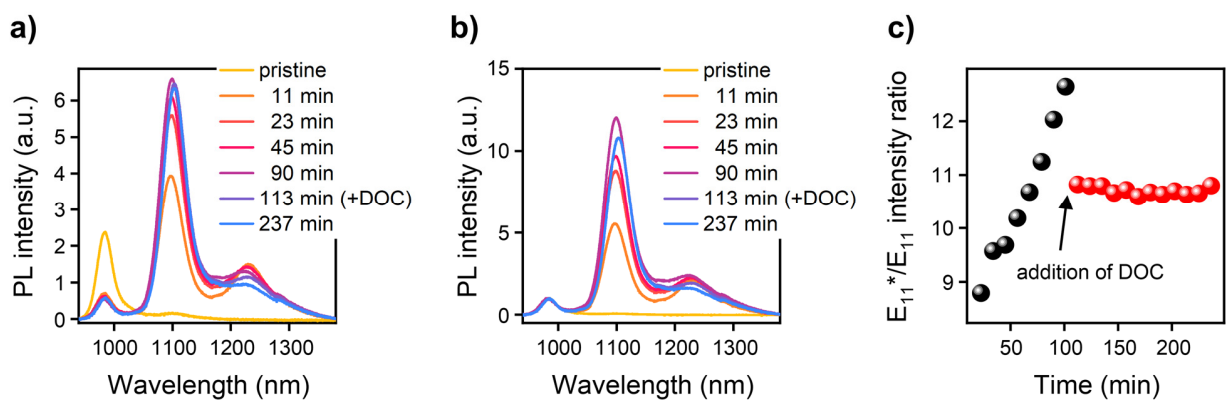

**Figure S8.** Controlling the defect density by surfactant exchange. The covalent functionalization of (6,5) SWNTs *via* Fenton-like reaction can be stopped by increasing the DOC concentration in the system to, *e.g.*, 0.2 % (w/v) after 113 min. Absolute (a) and normalized (b) PL spectra measured at various reaction times before and after the addition of DOC. The slight drop in $E_{11}^{*-}$ intensity (at ~ 1220 nm) originates from the reorganization of defect sites induced by excitation at the $E_{22}$ transition of (6,5) SWNTs. See Figures **S11-S13** and main text for a more detailed discussion. c) Evolution of $E_{11}^*/E_{11}$ PL intensity ratio with the reaction time. PL ratios extracted from PL spectra measured after the addition of DOC are indicated in red. The sudden change in PL ratio (by approximately 15%) originates from the fact, that surfactant exchange to DOC shows a larger impact on $E_{11}$ intensity (increased emission from mobile excitons) than $E_{11}^*$ intensity (localized defect emission).

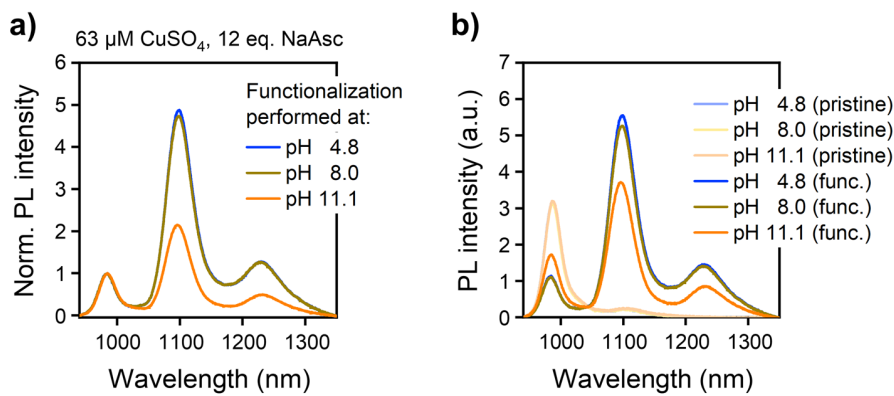

**Figure S9.** Impact of pH on the covalent functionalization of (6,5) SWNTs *via* Fenton-like reaction. Normalized (a) and absolute (b) PL spectra of SDS-coated (6,5) SWNTs when functionalized at pH 4.8, 8.0 and 11.1, respectively. The pH value shows a negligible effect on the PL properties of pristine (6,5) SWNTs. At a high pH value of ~11.1 a reduced reactivity is observed and can be explained by the formation of insoluble $Cu(OH)_2$. For lower pH values we expect the protonation of the surfactant to become the dominant factor as well as potential *p*-doping of the SWNTs.

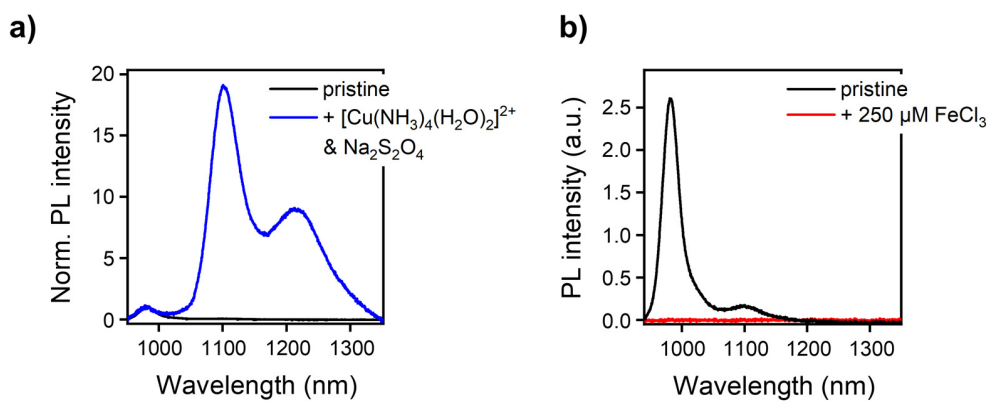

**Figure S10.** Functionalization *via* Fenton-like reaction with other reducing agents and metal catalysts. a) Normalized PL spectra of SDS-coated (6,5) SWNTs functionalized in the presence of 63 µM of $CuSO_4$ that was pre-treated with excess of ammonia and addition of 12 eq. sodium dithionite ($Na_2S_2O_4$). The presence of ammonia was crucial to prohibit the disproportionation of $Na_2S_2O_4$. Moreover, the formation of insoluble $Cu(OH)_2$ was prevented. b) PL spectra upon addition of 250 µM $FeCl_3$ to SDS-coated (6,5) SWNTs. The presence of Fe(III) ions is expected to lead to strong *p*-doping and thus PL quenching is observed.

**Reorganization of oxygen defects**

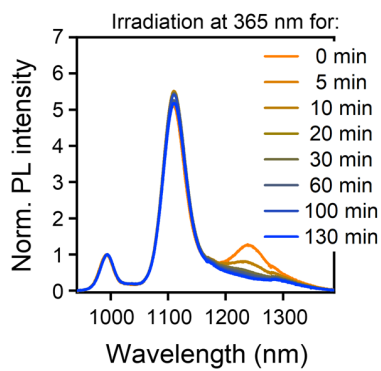

**Figure S11.** Normalized PL spectra of functionalized and subsequently DOC-coated (6,5) SWNTs after UV-light irradiation. The $E_{11}^*/E_{11}$ intensity ratios remain stable while the $E_{11}^{*-}/E_{11}$ intensity ratios decrease.

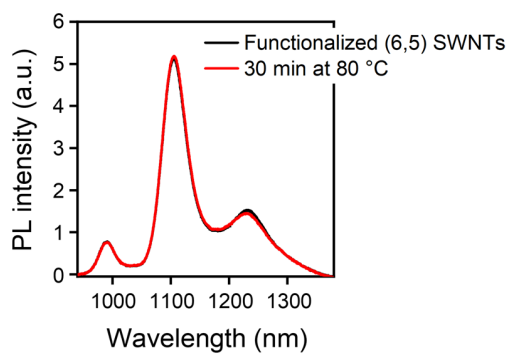

**Figure S12.** PL spectra of functionalized and subsequently DOC-coated (6,5) SWNTs heated to 80 °C for 30 min under exclusion of light. Note, that the small drop of $E_{11}^{*-}$ intensity and increase of $E_{11}^*$ intensity might occur already due to $E_{22}$ excitation during the PL measurement.

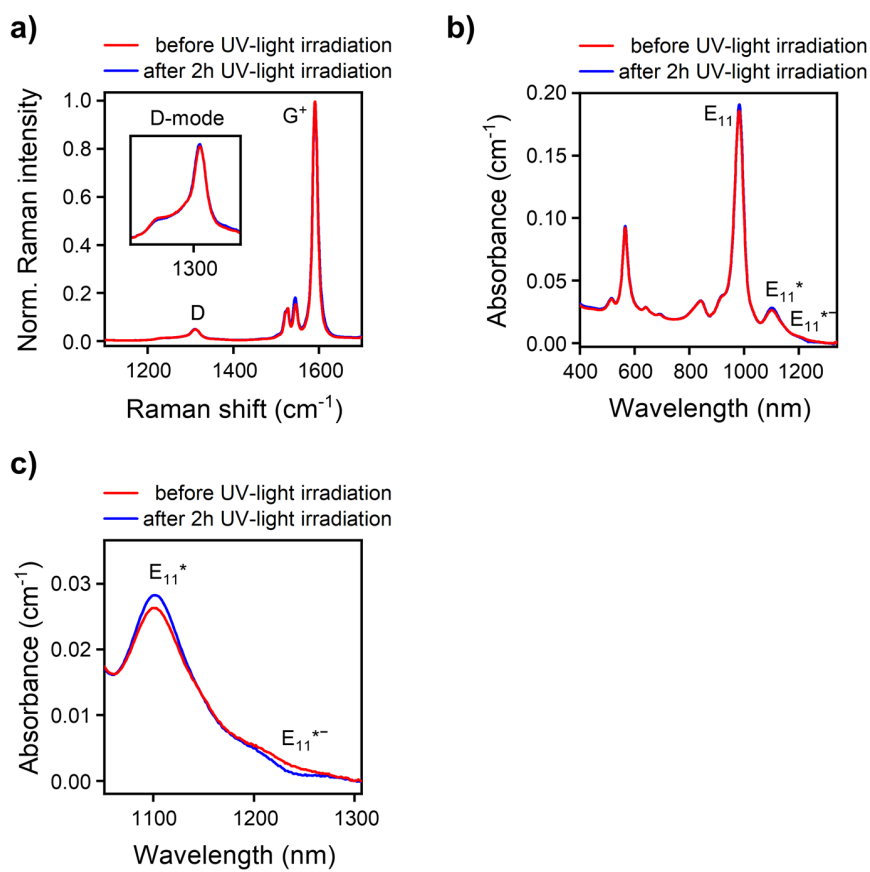

**Figure S13.** Effect of UV-light irradiation on Raman and absorption spectra. a) Raman spectra of functionalized DOC-coated (6,5) SWNTs before and after UV-light irradiation (365 nm) for 2 h. b) Absorption spectra of functionalized DOC-coated (6,5) SWNTs before and after UV-light irradiation (365 nm) for 2 h. c) Zoom-in on the $E_{11}*$ and $E_{11}*^-$ absorption band.

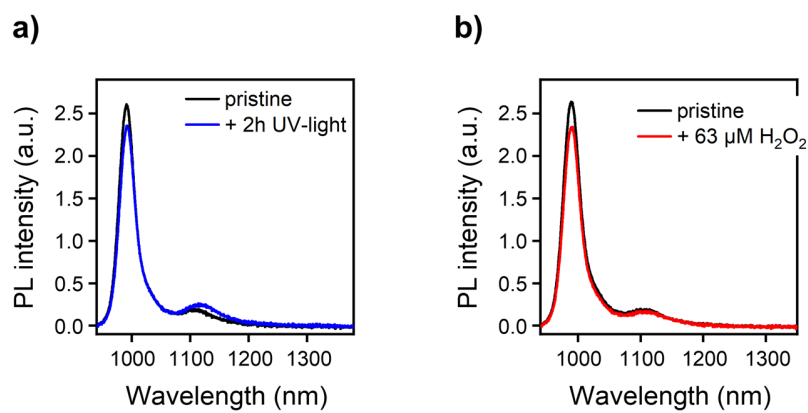

**Figure S14.** Investigation of potential PL quenching effects. a) PL spectra of SDS-coated (6,5) SWNTs before and after 2 h of UV-light (365 nm) irradiation. The slight decrease in $E_{11}$ emission and increase in PL intensity at ~1130 nm may indicate the formation of defects due to covalent functionalization induced by prolonged UV excitation of the SWNT. b) PL spectra of SDS-coated (6,5) SWNTs before and after addition of 63 μM $H_2O_2$.

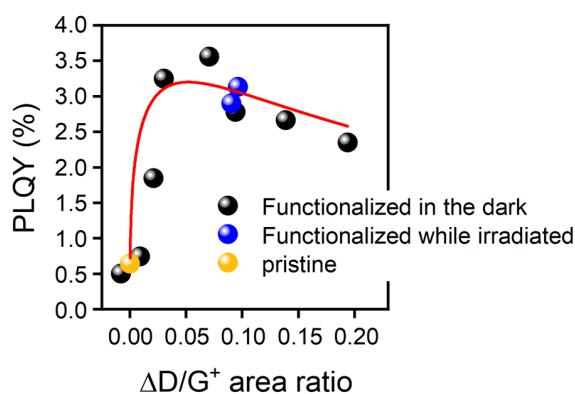

**Figure S15.** Extracted PLQYs of functionalized (6,5) SWNTs (after work-up) *vs.* integrated $\Delta D/G^+$ area ratio as a metric for the defect density (red line as guide to the eye). Measured PLQYs obtained by covalent functionalization under exclusion of light and under $E_{22}$ excitation are indicated in black and blue, respectively. The PLQY of pristine (6,5) SWNTs (yellow) represents an average value obtained from multiple separate measurements.

**Impact of the choice of surfactant on the functionalization process**

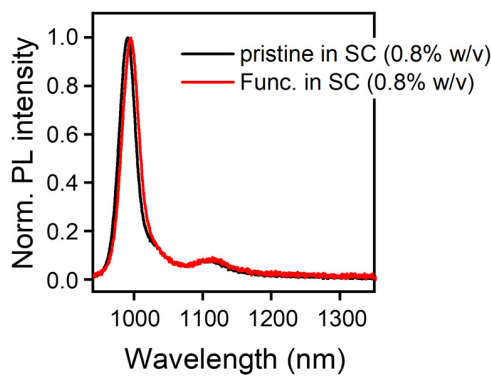

**Figure S16.** PL spectra measured before (black) and after functionalization with 63 µM CuSO$_4$ and 12 eq. NaAsc. The functionalization was conducted in aqueous SC dispersion at a concentration of 0.8 % (w/v), *i.e.* above the critical micelle concentration of SC.

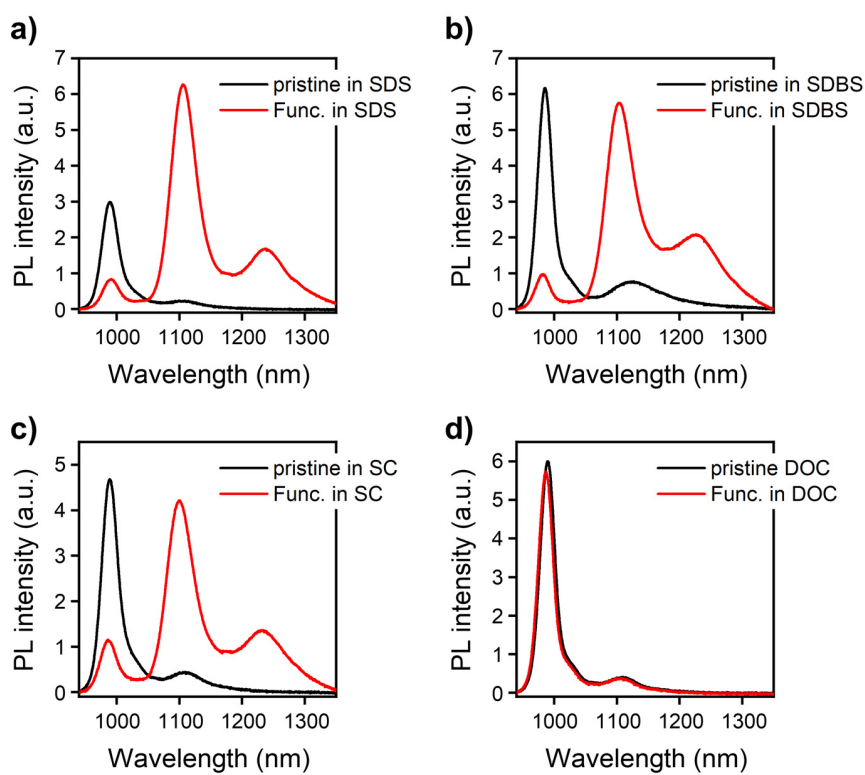

**Figure S17.** Impact of the surfactant on the covalent functionalization *via* Fenton-like reaction. PL spectra measured before (black) and after functionalization with 63 µM CuSO$_4$ and 12 eq. NaAsc. The functionalization was conducted in aqueous dispersions with SDS (a), SDBS (b), SC (c) and DOC (d) at a concentration of 0.33 % (w/v), *i.e.* above the critical micelle concentration for SDS, SDBS and DOC but below the critical micelle concentration for SC (see **Figure S16** for reference).

**Selection, characterization and functionalization of (7,5), (8,3) and (6,4) SWNTs**

**Table S2.** Exact volumes of added HCl and NaOCl used for the separation of (6,4), (8,3) and (7,5) SWNTs. For the separation of (6,4) SWNTs, the top phases after step 1 and step 2 were discarded and replaced by mimic suspensions. Separated (6,4) SWNTs partition to the top phase in step 3. For the separation of (8,3) and (7,5) SWNTs, the top phases in step 1 and bottom phases in step 2 were discarded and replaced by mimic suspensions. Separated (8,3) and (7,5) SWNTs partition to the bottom phase in step 3.

| Separated Species | Step 1 | Step 2 | Step 3 |
|---|---|---|---|
| **(6,4) SWNTs** | 10 µL HCl (0.5 M) | 20 µL HCl (0.5 M) | 40 µL HCl (0.5 M) 20 µL NaOCl |
| **(8,3) SWNTs** | 4 µL HCl (0.5 M) | 6 µL HCl (0.25 M) | 5 µL HCl (0.5 M) 10 µL NaOCl |
| **(7,5) SWNTs** | 8.5 µL HCl (0.5 M) | 2.5 µL HCl (0.25 M) | 5 µL HCl (0.5 M) 10 µL NaOCl |

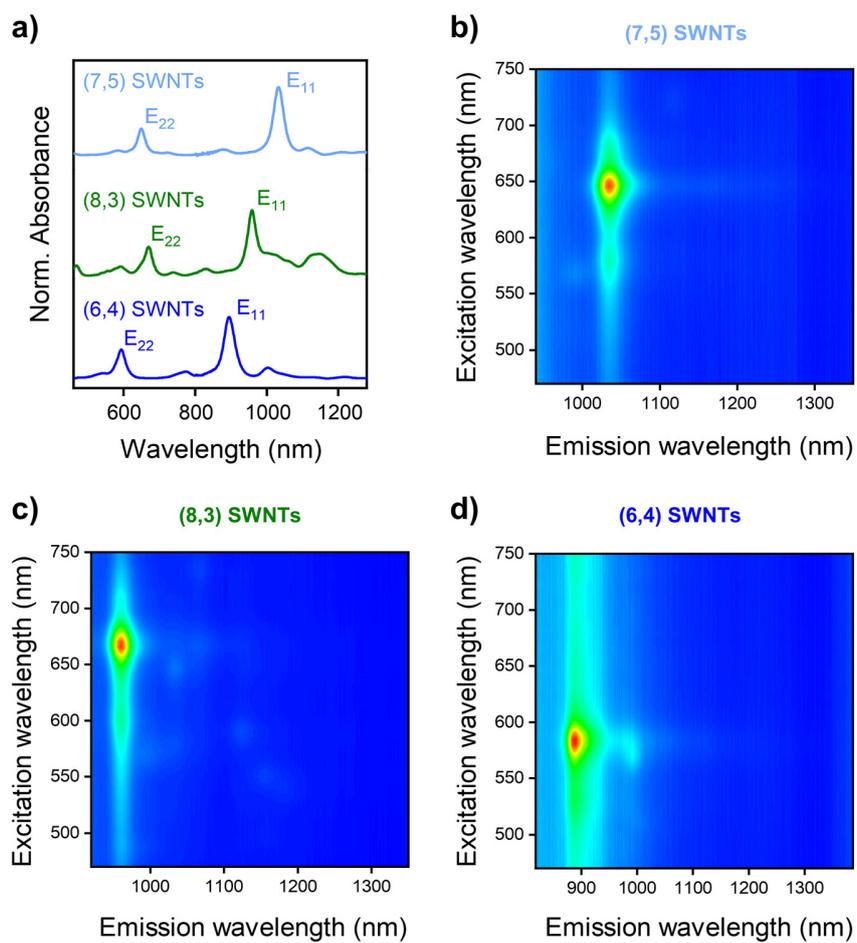

**Figure S18.** Spectroscopic characterization of ATPE sorted (7,5), (8,3) and (6,4) SWNTs dispersed in SDS. a) Absorption spectra of sorted (7,5), (8,3) and (6,4) SWNTs. b) PLE map of sorted (7,5) SWNTs. c) PLE map of sorted (8,3) SWNTs. d) PLE map of sorted (6,4) SWNTs.

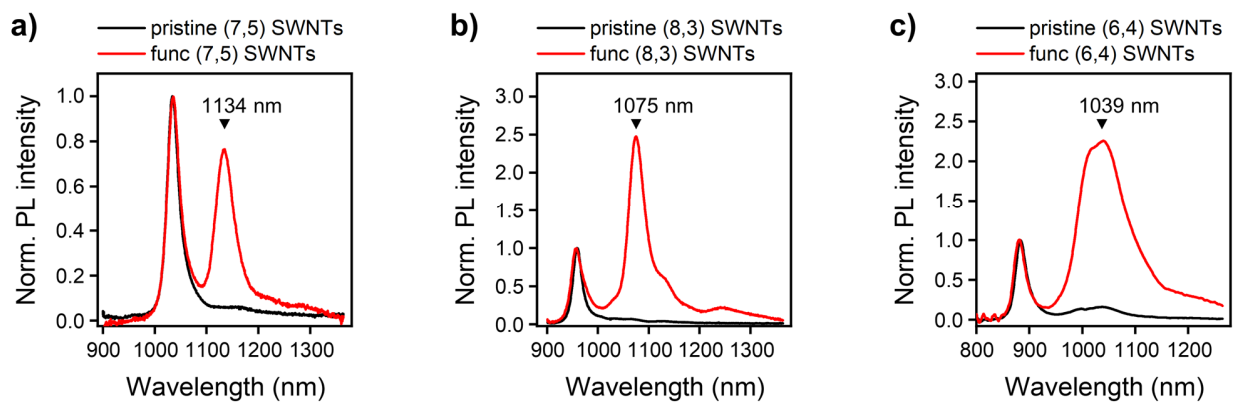

**Figure S19.** Covalent functionalization of (7,5), (8,3) and (6,4) SWNTs *via* Fenton-like reaction. Normalized PL spectra of SDS-coated (7,5) SWNTs (a), (8,3) SWNTs (b) and (6,4) SWNTs (c) functionalized with 63 µM $CuSO_4$ and 12 eq. NaAsc. All samples were exposed to light for several days before optical characterization. The defect state emission is blue-shifted compared to values reported by Ghosh *et al.*,[2] however, the exact peak position is expected to depend on the defect density.

**Table S3.** Optical trap depths and PLQYs of functionalized (7,5), (8,3) and (6,4) SWNTs.

| Sample (*d* - SWNT diameter) | Optical trap depth (meV) | $E_{11}$ share of PLQY (%) | $E_{11}^*$/PSB share of PLQY (%) | $PLQY_{total}$ (%) |
|---|---|---|---|---|
| **(7,5) SWNTs** (*d* = 0.829 nm) | - | 0.21 | 0.03 | 0.24 |
| **Func-(7,5) SWNTs** | 106 | 0.11 | 0.17 | 0.28 |
| **(8,3) SWNTs** (*d* = 0.782 nm) | - | 0.19 | 0.08 | 0.27 |
| **Func-(8,3) SWNTs** | 140 | 0.07 | 0.41 | 0.48 |
| **(6,4) SWNTs** (*d* = 0.692 nm) | - | 0.22 | 0.17 | 0.39 |
| **Func-(6,4) SWNTs** | 211 | 0.08 | 0.70 | 0.78 |

## Optical characterization and functionalization of unsorted CoMoCAT raw material

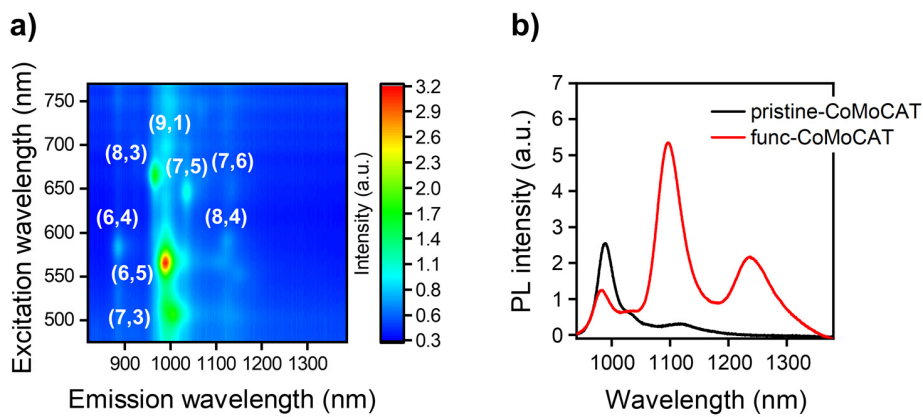

**Figure S20.** Covalent functionalization of unsorted CoMoCAT raw material *via* Fenton-like reaction. a) PLE map of pristine unsorted CoMoCAT raw material with identified SWNT chiralities. b) PL spectra obtained upon excitation at 570 nm of pristine and functionalized CoMoCAT raw material in 0.33 % (w/v) SDS.

## White light and NIR PL images of pristine and functionalized (6,5) SWNTs

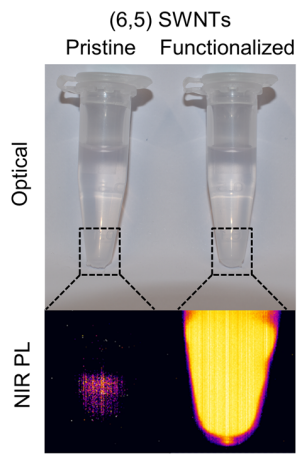

**Figure S21.** White light images (top) and NIR PL images (950-1600 nm) of pristine and functionalized dispersions of (6,5) SWNTs. NIR PL images were measured under identical conditions upon excitation at 785 nm at equal concentrations. The displayed insets represent the areas that are shown in the NIR PL image.

**Length distribution of shortened functionalized (6,5) SWNTs**

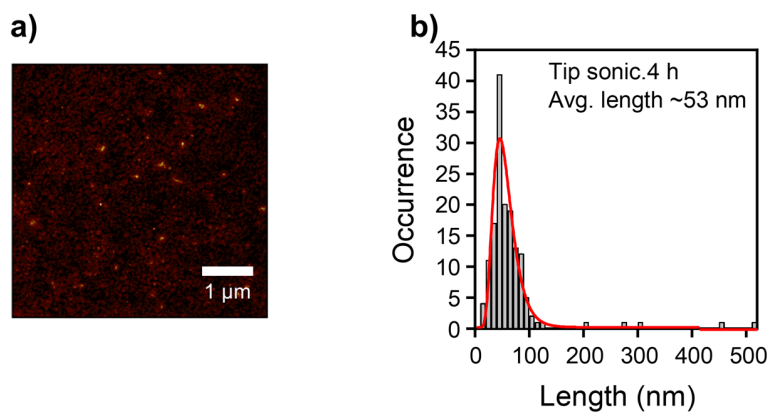

**Figure S22.** Length distribution of shortened functionalized (6,5) SWNTs. a) Atomic force micrograph (5x5 μm$^2$) of tip-sonicated (4 hours) (6,5) SWNTs after functionalization *via* Fenton-like reaction. b) SWNT length histogram fitted by a log-normal distribution with an average length of 53 nm.

**Surfactant transfer to ssDNA**

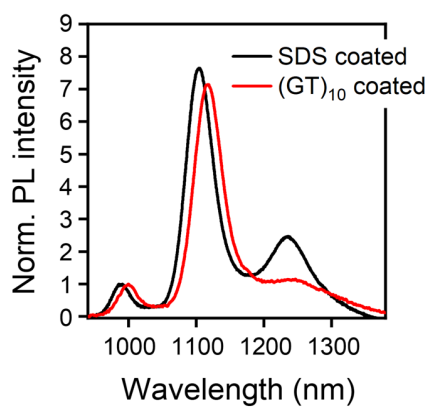

**Figure S23.** Surfactant exchange with ssDNA. Normalized PL spectra of functionalized (6,5) SWNTs before surfactant transfer (SDS, black) and after transfer to ssDNA-(GT)$_{10}$. E$_{11}$ and E$_{11}$* emission are red-shifted for 11 and 13 nm, respectively, after transfer to ssDNA-(GT)$_{10}$.

## Determination of copper concentration with ICP-OES

**Table S4**. Copper ion concentration obtained by ICP-OES. Concentrations of (6,5) SWNTs are based on the SWNT concentration used for functionalization. SWNT concentrations were determined based on fitted absorption spectra.[3, 4]

| Name | Sample | Copper conc. ($\mu g\ L^{-1}$) | Average copper conc. ($\mu g\ L^{-1}$) | (6,5) SWNT conc. ($mg\ L^{-1}$) |
|---|---|---|---|---|
| Reference | 1 | 4.2 | 4.0 | - |
| | 2 | 3.8 | | |
| Pristine (6,5) SWNTs | 1 | 5.9 | 5.2 | 0.6 |
| | 2 | 4.4 | | |
| Functionalized (6,5) SWNTs | 1 | 6.6 | 51.2 | 0.6 |
| | 2 | 95.8 | | |
| Pristine CoMoCAT SWNTs | 1 | 7.9 | 6.8 | 0.6 |
| | 2 | 5.7 | | |
| Functionalized CoMoCAT SWNTs | 1 | 73.5 | 59.3 | 0.6 |
| | 2 | 45 | | |